\documentclass{iopjournal}
\usepackage{times}
\usepackage[numbers,sort&compress]{natbib}

\usepackage{amsmath,amsfonts,bm}

\def\1{\bm{1}}

\DeclareMathAlphabet{\mathsfit}{\encodingdefault}{\sfdefault}{m}{sl}
\SetMathAlphabet{\mathsfit}{bold}{\encodingdefault}{\sfdefault}{bx}{n}

\usepackage{hyperref}
\usepackage{url}
\usepackage{xcolor}
\usepackage{tcolorbox}
\usepackage{tikz}
\usepackage{cleveref}
\usepackage{booktabs}
\usepackage{array}
\usetikzlibrary{arrows.meta}
\usepackage{subcaption}
\usepackage{bm}

\hypersetup{colorlinks=true, linkcolor=black, citecolor=black, urlcolor=blue}

\renewcommand{\headrulewidth}{0pt}
\renewcommand{\articletype}[1]{}

\newcommand{\code}[1]{\texttt{#1}}

\begin{document}

\articletype{Paper}

\title{Real-time Tokamak Equilibrium Reconstruction Under Limited Experimental Data via Physics-Grounded Synthetic Pre-training}

\author{Mingdong He$^{1,\dagger,\ddagger}$, Qixian Hu$^{2,\dagger,\ddagger}$, Tengfei Xu$^{3,4}$, Tailin Wu$^{4}$ and Zheng Xiong$^{3,*}$}

\affil{$^{1}$Mathematical Institute, University of Oxford, Oxford, United Kingdom}

\affil{$^{2}$School of Engineering Science, University of Chinese Academy of Sciences, Beijing, China}

\affil{$^{3}$Shanghai Innovation Institute, Shanghai, China}

\affil{$^{4}$Westlake University, Hangzhou, China}

\affil{$^{\dagger}$Mingdong He and Qixian Hu contributed equally to this work.}

\affil{$^{*}$Author to whom any correspondence should be addressed.}

\email{\href{mailto:mingdong.he@maths.ox.ac.uk}{mingdong.he@maths.ox.ac.uk}; \href{mailto:xiongzheng@sii.edu.cn}{xiongzheng@sii.edu.cn}}


\begin{abstract}
Real-time equilibrium reconstruction is essential for tokamak plasma control. Data-driven surrogates are usually trained on high-quality labeled experimental data, but producing a large amount of labeled data sufficient for surrogate learning requires many expensive discharge shots, and few exist at the start of a new campaign or the operation of a new device. To reduce this cost, we propose to pre-train on simulated equilibria, generated at low cost by physics-based simulation, then fine-tune on limited real data. We evaluate under three settings with increasing difficulties, i.e., in-distribution, out-of-distribution (shape), and cross-campaign (temporal) splits, which mirror how reconstruction is actually deployed. 
Experimental results validate that synthetic pre-training is highly effective when few real labels are available: fine-tuning on only 1\% of the full available real dataset cuts the per-sample normalized root-mean-square error (nRMSE) of the reconstructed poloidal flux by roughly 58\% under out-of-distribution shapes and 64\% under cross-campaign extrapolation, compared to a model trained from scratch on the same amount of real data. These realistic evaluations show that synthetic pre-training improves equilibrium reconstruction accuracy in the data-scarce regime that real tokamak campaigns face.
\end{abstract}

\section{Introduction}

The plasma state inside a tokamak is not directly observable and must be inferred from external magnetic diagnostics, a task known as equilibrium reconstruction. The standard solver, the Equilibrium Fitting (EFIT) code~\citep{lao1985reconstruction}, iteratively solves the Grad--Shafranov equation to obtain a self-consistent equilibrium that matches the measurements. However, its computational cost and dependence on suitable initialization hinder real-time deployment. Supervised data-driven surrogates offer a route to fast inference~\citep{joung2020deep,wai2022neural,stokolesov2025neural}. Existing methods trained on reconstructed experimental equilibria have demonstrated promising performance~\citep{zheng2024efitnn,madireddy_efit-prime_2024,mcclenaghan_augmenting_2024,xia_reconstruction_2026}, but require large quantities of high-quality labeled data. Such labels are costly to obtain and are particularly scarce at the beginning of a new experimental campaign, when few equilibria from that campaign have been reconstructed.

Simulation provides a scalable way to alleviate this label scarcity. Previous studies have shown that surrogate models can be trained on synthetic Grad--Shafranov solutions, including neural operators~\citep{ding2026physicsinformedicml,bonotto_reconstruction_2024}, networks trained on MCMC-generated equilibrium libraries~\citep{agnello_emulation_2024}, and other synthetic-data surrogates~\citep{maurya2026deeplearningadityau}. These studies demonstrate the feasibility of learning numerical Grad--Shafranov solution maps from synthetic equilibria. However, magnetic-diagnostic equilibrium reconstruction uses real measurements, which can differ from simulated diagnostics because of measurement noise and device-specific effects. Hu et al.\cite{realpdebench2026} shows that models trained on simulated data generalize less effectively than models trained on real data when both are evaluated on the same real-world test sets. Consequently, strong performance on synthetic test data alone does not establish accuracy on real discharges. 

In this paper, we therefore propose a two-stage surrogate learning method consisted of synthetic pre-training followed by fine-tuning on a limited amount of real data. Synthetic pre-training provides abundant physics-grounded supervision, while real-data fine-tuning adapts the model to experimental measurements and device conditions. This strategy has improved performance and convergence in real-world PDE benchmarks~\citep{realpdebench2026} and improved data efficiency in plasma-diagnostic analysis when experimental samples are limited~\citep{van_hoomissen_transfer_2026}. We develop a sim-to-real framework for reconstructing plasma equilibria directly from tokamak diagnostic measurements and systematically evaluate the benefits of synthetic pre-training across different real-data fractions on MAST~\citep{jackson2024fairmast}.


To this end, we construct an end-to-end sim-to-real pipeline for free-boundary Grad--Shafranov equilibrium reconstruction. Synthetic equilibria are generated by perturbing the profile coefficients along the leading principal components anchored to real EFIT statistics, solving the free-boundary problem with the FreeGSNKE solver \cite{amorisco2024freegsnke}, and filtering the solutions using convergence and physical-consistency criteria. This procedure yields 213{,}924 aligned synthetic input--output pairs without manual labelling.

We systematically evaluate synthetic pre-training under three protocols of increasing difficulty and practical relevance: an in-distribution random split, a shape out-of-distribution (Shape-OOD) split, and a cross-campaign temporal-OOD split. Under the two OOD generalization settings, the benefit is persistent: pre-training improves accuracy at every real-data fraction, with larger gains as real data become scarcer, reaching an nRMSE reduction of approximately 64\% with only 1\% real data under cross-campaign extrapolation.

These results are of practical significance. The synthetic generation pipeline requires no manual labelling effort and is directly transferable to other tokamaks and operational regimes where high-quality real equilibrium labels are scarce. The three-protocol evaluation demonstrates that synthetic pre-training improves not only in-distribution accuracy but also generalization to unseen plasma shapes and later experimental campaigns, while the reconstructed flux is shown to be consistent with independent measurements of the total plasma current. Our quantitative results of the data-efficiency indicate how much real experimental data can be saved to reach a given reconstruction accuracy, mitigating the scarcity of high-quality real equilibrium labels that has long been the main bottleneck of data-driven equilibrium reconstruction and bringing fast surrogate models closer to real-time tokamak control.

The remainder of this paper is organized as follows. \Cref{sec:data} describes the MAST experimental data and the selection of the real equilibrium slices used for training. \Cref{sec:gen} explains how synthetic equilibria are generated, covering profile parameterization, statistical perturbation, forward solving, and physical screening. \Cref{sec:method} formulates equilibrium reconstruction as a supervised regression problem, introduces the reconstruction model, and specifies the two-stage simulation pre-training plus real fine-tuning protocol. \Cref{sec:experimental_design} introduces the experimental settings, including the data budget, generalization settings, and evaluation metrics. \Cref{sec:results} presents the experimental results and evaluates the computational efficiency of the proposed model. \Cref{sec:conclusion} concludes the paper.

\section{Data preparation}
\label{sec:data}
The FAIR-MAST dataset~\cite{jackson2024fairmast} provides open-access data for \(11{,}573\) MAST discharges across the M5--M9 campaigns, spanning wall geometry, magnetics, coil currents, and EFIT reconstruction results. We denote the diagnostic signals by $\mathbf{x}_{\text{real}}$. Their EFIT reconstruction provides the poloidal flux $\psi(R,Z)$ and derived quantities including the pressure and current-profile gradients $p'$ and $FF'$, the last closed flux surface (LCFS), the magnetic axis, and global parameters such as the plasma current $I_p$, the poloidal beta $\beta_{\mathrm{pol}}$, the internal inductance $l_i$, and the safety factor at the 95\% normalized poloidal flux surface $q_{95}$.

\begin{figure}[t]
    \centering
    \includegraphics[width=0.6\linewidth]{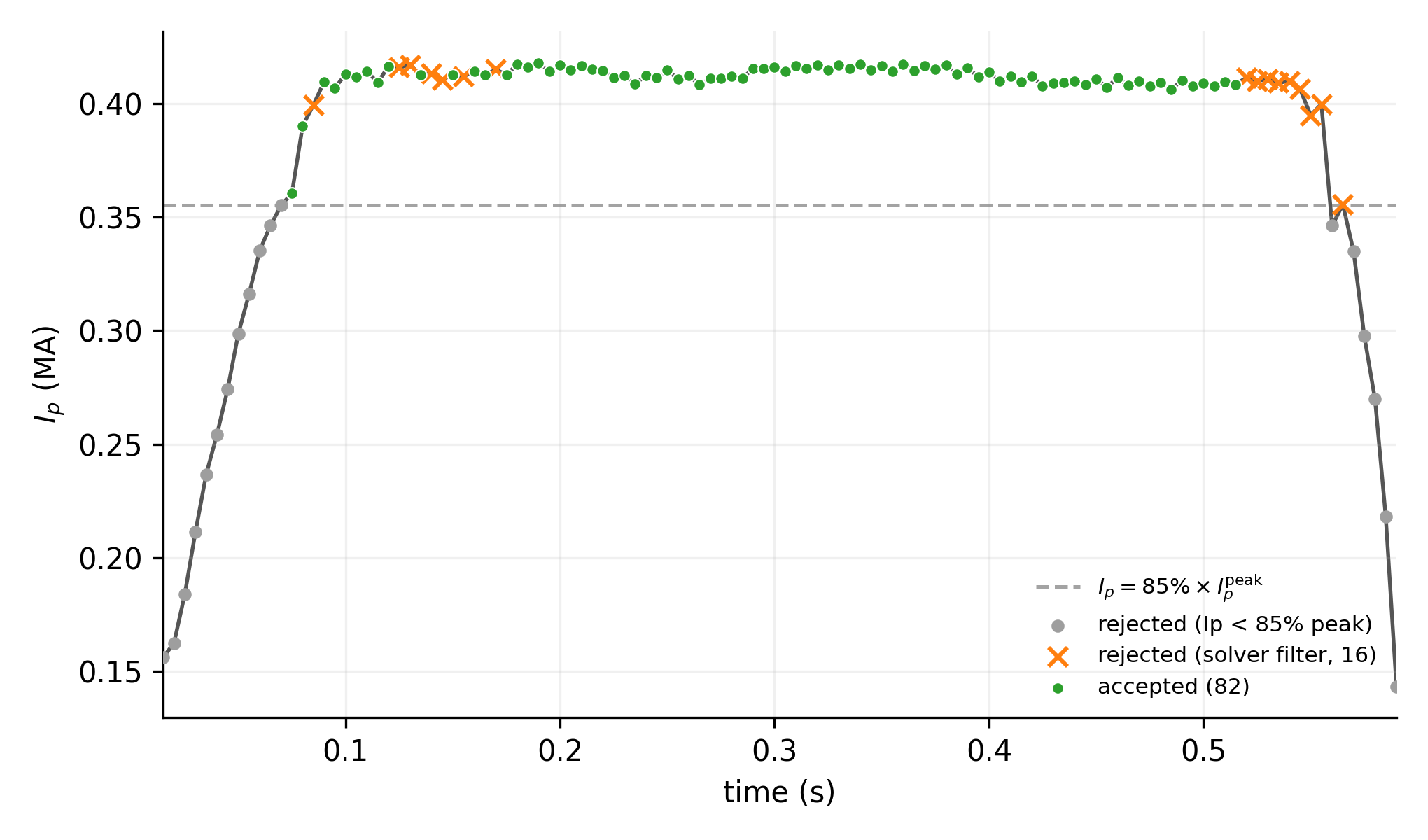}
    \caption{Quality of the screened samples. Filtered slices of example shot 20436 along the full time axis: retained frames (green) are concentrated in the flat-top region where the plasma current is near its peak and the EFIT labels are most reliable.}
    \label{fig:ip_screening}
\end{figure}

Raw MAST discharges are not directly usable for training, because many lack complete diagnostics or EFIT reconstructions, and the labels are reliable only within the flat-top phase \cite{pentland2025validation}. We therefore screen the data in four steps. First, we retain only discharges with complete diagnostics and EFIT reconstructions. Second, we restrict each shot to its flat-top phase, keeping frames where $I_p$ is above $85\%$ of the shot peak, since the plasma is then quasi-stationary and the EFIT labels are most reliable. Third, we generate one paired synthetic equilibrium per retained slice, as described in \Cref{sec:gen}, so that every real training example has a matching simulation. Finally, we keep only converged, physically consistent solutions. \Cref{fig:ip_screening} confirms that the retained frames concentrate in the flat-top window, and \Cref{app:data_prep} summarises the full cascade, which reduces the $11{,}573$ raw discharges to $213{,}924$ real slices from $7{,}286$ discharges, each matched to a synthetic equilibrium and used for training and evaluation in \Cref{sec:method}.

\section{Generation of simulation data}
\label{sec:gen}

\subsection{The Grad--Shafranov equation}
The Grad--Shafranov equation (GSE)~\cite{grad1958hydromagnetic}
describes axisymmetric tokamak equilibria. It arises from the ideal MHD
force balance $\nabla p=\mathbf{J}\times\mathbf{B}$: under axisymmetry,
this three-dimensional vector equation reduces to a two-dimensional scalar
partial differential equation for the poloidal flux $\psi(R,Z)$ in
cylindrical coordinates $(R,\phi,Z)$,
\begin{equation}
  \Delta^{*}\psi=-\mu_{0}RJ_{\phi}(\psi),
  \label{eq:gs}
\end{equation}
where the Grad--Shafranov operator is
$\Delta^{*}=R\partial_{R}(R^{-1}\partial_{R})+\partial_{Z}^{2}$, and the
toroidal current density $J_{\phi}(\psi)$ acts as a nonlinear source term,
\begin{equation}
  J_{\phi}(\psi)=R\frac{dp(\psi)}{d\psi}
  +\frac{F(\psi)}{\mu_{0}R}\frac{dF(\psi)}{d\psi},
  \label{eq:jphi_psi}
\end{equation}
determined by the pressure $p(\psi)$ and poloidal current $F(\psi)$
profiles, where $\mu_0$ is the magnetic permeability of free space. The dependence of the $\psi$ makes the GSE a nonlinear elliptic PDE, which is typically solved numerically with iterative methods. The solution $\psi(R,Z)$ determines the magnetic geometry, including the LCFS, the magnetic axis, and the plasma shape and position.

\subsection{Free-boundary solve}
In practical equilibrium reconstruction, the plasma boundary is generally not known a priori and must instead be determined self-consistently from diagnostic constraints, motivating a free-boundary formulation of the GSE~\cite{lao1985reconstruction}. Free-boundary solvers seek the solution over a computational domain
that extends beyond the plasma, with the poloidal flux on the outer
boundary determined by the currents in external poloidal-field coils
through the free-space Green's function of the Grad--Shafranov operator.
The plasma--vacuum interface, i.e.~the last closed flux surface or the
separatrix in diverted configurations, is then identified as part of the
solution, typically by a double-nested Picard iterative scheme in which an
inner fixed-boundary solve is alternated with an outer loop that updates the
coil currents and the boundary flux condition. This formulation allows
desired equilibrium properties such as plasma shape, position, and total
current to be prescribed, and is therefore essential for equilibrium
reconstruction, scenario design, and the initialization of time-dependent
MHD simulations. 

In this work we solve the free-boundary problem with the
FreeGSNKE solver \cite{amorisco2024freegsnke}, which requires the following source terms
\begin{equation}
\theta_{\mathrm{GS}}=\{p',\,FF',\,I_p,\,F_{\mathrm{vac}},\,I_{\mathrm{PF}},\,I_{\mathrm{passive}},\, \text{machine geometry}\}  
\end{equation}
to obtain the simulation data $\psi_{\text{sim}}$. Here, $p'=\mathrm{d}p/\mathrm{d}\psi$ is the pressure-gradient profile, and $FF'=F\,\mathrm{d}F/\mathrm{d}\psi$ is the toroidal-field source profile, where $F=RB_\phi$ is the toroidal magnetic field profile and $B_{\phi}$ is the toroidal component of the magnetic field. Furthermore, $I_p$ is the total plasma current, $F_{\mathrm{vac}}=R_{\mathrm{mag}}B_{\mathrm{vac}}$ is the vacuum toroidal-field function, $I_{\mathrm{PF}}$ and $I_{\mathrm{passive}}$ denote the currents in the active poloidal-field coils and passive conducting structures, respectively, and the machine geometry specifies the spatial configuration of the coils, passive structures, and limiter. The machine geometry is fixed and $I_{\mathrm{PF}}$ and $I_{\mathrm{passive}}$ are taken from measured coil currents. The detailed source terms and machine geometry are given in \Cref{tab:gs_inputs,tab:machine_geometry} in Appendix \ref{app:source_terms}. 

All source profiles are expressed as functions of the normalized poloidal flux
\begin{equation}
  \overline{\psi} = \frac{\psi_{\mathrm{axis}}-\psi}{\psi_{\mathrm{axis}}-\psi_{\mathrm{boundary}}} \in [0, 1],
\label{eq:norm_flux}
\end{equation}
where $\psi_{\mathrm{axis}}$ and $\psi_{\mathrm{boundary}}$ are the poloidal fluxes at the magnetic axis and at the LCFS, respectively, which are both determined self-consistently by the free-boundary solve; thus $\overline{\psi}=0$ at the magnetic axis and $\overline{\psi}=1$ at the LCFS.

FreeGSNKE solves the free-boundary problem by iterating two coupled steps, and its static solver has been validated against the Fiesta equilibrium solver \cite{cunningham2013high} on EFIT++ reconstructions from MAST-U, confirming the effectiveness and accuracy of its free-boundary solutions \cite{pentland2025validation}. First, the toroidal current density is evaluated from the fitted profiles evaluated at $\overline{\psi}$,
\begin{equation}
J_{\phi}(R,Z) = L\left[\,R\,p'(\overline{\psi}) + \frac{1}{\mu_0 R}\,FF'(\overline{\psi})\right],
\label{eq:jphi}
\end{equation}
where $L = I_p / \int J_{\phi}\,dRdZ$ renormalizes the total current to the target $I_p$, with $J_{\phi}$ set to zero outside the last closed flux surface (LCFS); plugging \eqref{eq:jphi} into $\Delta^{*}\psi=-\mu_0 R J_{\phi}$ recovers the GS equation \eqref{eq:gs} up to this normalization. Second, the plasma current is mapped to the poloidal flux on the computational boundary by a Green's-function convolution, $\psi|_{\partial\Omega}=\int_S G(R,Z;R',Z')\,J_{\phi}(R',Z')\,dR'dZ'$, using the analytic flux kernel of a toroidal current filament; the interior flux is then obtained by solving the linear elliptic problem $\Delta^{*}\psi=-\mu_0 R J_{\phi}$ with $\psi|_{\partial\Omega}$ as the Dirichlet boundary condition. Iterating these two steps with a Newton--Krylov scheme until the GS residual reach to a certain tolerance determines the magnetic axis and the LCFS self-consistently.

\begin{figure}[t]
\centering
\includegraphics[width=0.9\textwidth]{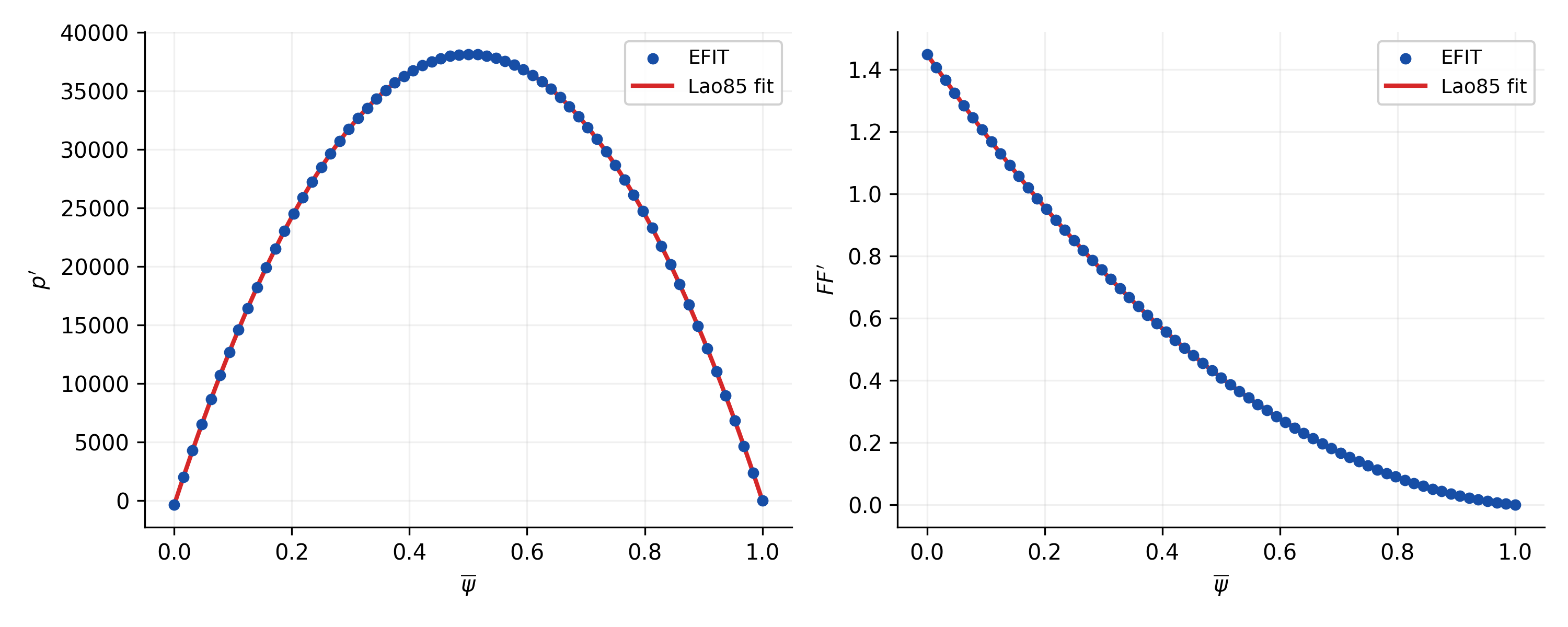}
\caption{Lao85 polynomial fits of the source profiles for MAST shot 11766 at $t=0.18$ s. Left: pressure gradient $p'(\psi)$ with coefficients $\alpha=(-377.63,\,1.5361\times 10^5,\,-1.5323\times 10^5)$, satisfying $p'(1)=0$ at the LCFS boundary. Right: toroidal-field source profile $FF'(\psi)$ with coefficients $\beta=(1.4492,\,-2.7103,\,1.2611)$. The horizontal axis is the normalized poloidal flux $\overline{\psi}$ ($0$ = magnetic axis, $1$ = LCFS).}
\label{fig:lao_profiles}
\end{figure}

To make our sampling easier, we do not rerun EFIT to obtain the profiles as those in \cite{pentland2025validation,pentland2025multiple}. Instead, we fit the profiles by using Lao's polynomials \cite{lao1985reconstruction} from the FAIR-MAST EFIT reconstructions,
\begin{align}
  p'(\overline{\psi})
    &= \sum_{i=0}^{n_p} \alpha_i \overline{\psi}^{\,i}
     - \left(\sum_{i=0}^{n_p} \alpha_i\right) \overline{\psi}^{\,n_p+1}, \\
  FF'(\overline{\psi})
    &= \sum_{i=0}^{n_F} \beta_i \overline{\psi}^{\,i}
     - \left(\sum_{i=0}^{n_F} \beta_i\right) \overline{\psi}^{\,n_F+1},
  \label{eq:lao_profiles}
\end{align}
where the profiles $p'$ and $FF'$ are taken to be zero at the boundary via the corrected basis, which corresponds to $\bar{\alpha}=\bar{\beta}=1$ in the general Lao85 parameterization. In this work, we choose $n_p = n_F = 2$. \Cref{fig:lao_profiles} shows the fitted profiles and associated coefficient values. 

Therefore, we could sample the following quantity 
\begin{equation}
\tilde{\theta}_{\mathrm{pert}}
=
\{\alpha_i,\beta_i,I_{\mathrm{PF}}\},
\end{equation}
while the passive-structure currents and machine geometry are inherited
from the corresponding experimental state. The perturbed parameters,
together with these fixed or inherited quantities, define the complete
input state for each FreeGSNKE solve.

To generate synthetic equilibria, we perturb selected parameters
of each clean-pool sample and solve the corresponding
free-boundary Grad--Shafranov problem. Each sample is described by a $16$-dimensional parameter vector $\mathbf{x}$, comprising the six Lao85 profile coefficients $\boldsymbol{\alpha}$ and $\boldsymbol{\beta}$ together with the ten active-coil currents. The plasma current $I_p$ and the vacuum toroidal-field function $F_{\mathrm{vac}}$ were held fixed at their measured values, and the zero-variance coils were excluded from the perturbation space.

We use PCA to perturb parameters along the principal directions of the real-data covariance, preserving the physical couplings among coils and profiles to avoid generating physically inconsistent independent combinations. We standardize the clean-pool samples and compute the eigendecomposition of their correlation matrix over the $502{,}402$ clean samples described in \Cref{app:data_prep}. Almost all of the variability is carried by a few directions, so we retain the first eight principal components, which capture around $98.5\%$ of the total variance, and add noise only within the subspace they span. Each sample was perturbed as
\begin{equation}
\mathbf{x}'=\mathbf{V}\bigl(\mathbf{t}+\boldsymbol{\delta}\bigr)\odot\boldsymbol{\sigma}+\boldsymbol{\mu},
\label{eq:pca_perturb}
\end{equation}
where $\mathbf{t}=\mathbf{V}^{\mathsf{T}}\mathbf{x}_{\mathrm{std}}$ are the principal-component scores and $\odot$ denotes elementwise multiplication, and the perturbation follows $\delta_k\sim\mathcal{U}(-0.1,\,0.1)\sqrt{\lambda_k}$ along each retained component. The amplitude of $\pm 0.1\sigma$ was chosen heuristically to introduce sufficient variability while keeping the perturbed samples close to the experimentally observed parameter distribution. After perturbation, parameter values outside the range between
their $1$st and $99$th empirical percentiles in the clean pool
are set to the nearest endpoint of that range. This limits
extreme parameter values in the synthetic samples. The boundary conditions $p'(1)=FF'(1)=0$ are enforced automatically by the corrected Lao85 basis, with the highest-order coefficient set to the negative sum of the free coefficients. Each sample was perturbed once, preserving one-to-one pairing with the real samples, and the resulting parameter sets were solved with the free-boundary Grad--Shafranov solver FreeGSNKE to obtain synthetic equilibria.

\Cref{tab:perturbation} summarizes the composition of the PCA perturbation space.

\begin{table}[htbp]
\centering
\small
\caption{Composition of the PCA perturbation space \eqref{eq:pca_perturb}.}
\label{tab:perturbation}
\begin{tabular}{lcl}
\toprule
Parameter group & No. & Action \\
\midrule
Lao85 profile coefficients
($a_0,a_1,a_2,b_0,b_1,b_2$)
& 6 & Perturbed \\

\shortstack[l]{Active-coil currents\\
(\texttt{P2IL}, \texttt{P2IU}, \texttt{P2OL}, \texttt{P2OU},\\
\texttt{P3L}, \texttt{P3U}, \texttt{P4L}, \texttt{P4U},\\
\texttt{P5L}, \texttt{P5U})}
& 10 & Perturbed \\

\addlinespace
Plasma current $I_p$
& 1 & Fixed ($\int J_{\phi}\,\mathrm{d}R\,\mathrm{d}Z=I_p$) \\

Vacuum toroidal-field function $F_{\mathrm{vac}}$
& 1 & Fixed (quasi-constant) \\

\addlinespace
$P_1$/$P_6$ coil currents
& 2 & Excluded (zero variance) \\
\bottomrule
\end{tabular}
\end{table}

\subsection{Synthetic diagnostics}
The generated magnetic poloidal flux $\psi_{\text{sim}}(R, Z)$ can be used to generate synthetic diagnostics so that it yields aligned synthetic pairs $(\mathbf{x}_{\text{sim}},\,\psi_{\text{sim}})$ in the same feature space as the real pairs $(\mathbf{x}_{\text{real}},\,\psi_{\text{real}})$.

For each accepted synthetic equilibrium, the corresponding magnetic diagnostic signals were generated through a forward model of the MAST diagnostics: flux-loop and pickup-coil signals are computed from the poloidal flux via the analytic Green's-function kernel of toroidal current filaments, evaluated at the probe and loop positions given by the machine geometry; coil currents are taken directly from the perturbed source parameters. This yields a 68-dimensional vector of synthetic diagnostic and
current features. Discharge time is additionally included to form
the 69-dimensional reconstruction input. 
Each synthetic sample retains the recorded discharge time of its
corresponding experimental seed. The discharge time serves as contextual
metadata for the locally perturbed equilibrium. Real diagnostic signals are noisy, whereas the forward-modelled ones are not, so we inject calibrated noise to bring the synthetic features closer to real ones. To study the robustness of synthetic pre-training to diagnostic noise, we construct three variants of the synthetic dataset that differ only in the noise level of the diagnostic features, while sharing identical $\psi$ labels and equilibria: \emph{Clean} (no noise), \emph{Noisy 0.5}, and \emph{Noisy 2}. In these two noisy variants, zero-mean Gaussian noise is added independently to each diagnostic channel:
\begin{equation}
x_{\mathrm{noisy},c}
=
x_{\mathrm{clean},c}
+
s\sigma_c\epsilon_c,
\qquad
\epsilon_c\sim\mathcal{N}(0,1),
\label{eq:noise}
\end{equation}
where \(s\in\{0.5,\,2.0\}\) is a global scaling factor and \(\sigma_c\) is the calibrated reference noise standard deviation for channel \(c\), as detailed in Appendix~\ref{app:noise_calibration}. Thus, \emph{Noisy 0.5} and \emph{Noisy 2} use \(0.5\times\) and \(2.0\times\) the reference noise level, respectively, while \emph{Clean} corresponds to \(s=0\). Noise is applied only to the diagnostic features and never to the \(\psi\) labels or equilibrium solutions, which are shared across all three variants. The time feature is kept unchanged across the three noise variants.

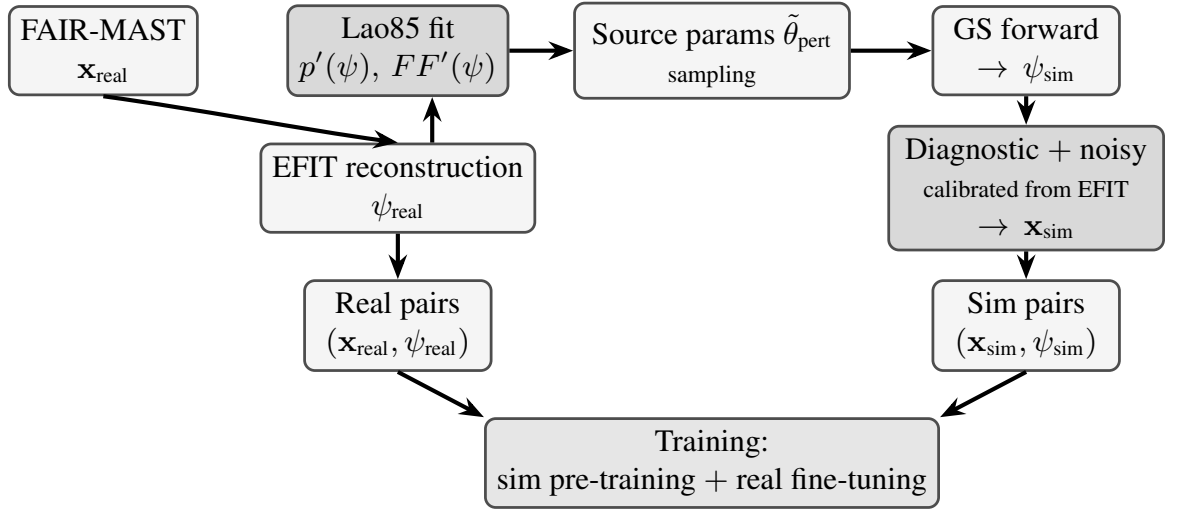
\begin{figure}[t]
\centering
\resizebox{\columnwidth}{!}{%
\begin{tikzpicture}[
  box/.style={draw=black!70, thick, rounded corners=3pt, fill=gray!8, align=center,
              minimum width=1.9cm, minimum height=0.9cm, font=\small},
  key/.style={draw=black!70, thick, rounded corners=3pt, fill=gray!20, align=center,
              minimum width=3.4cm, minimum height=0.9cm, font=\small},
  choice/.style={draw=black!70, thick, rounded corners=3pt, fill=gray!30, align=center,
              minimum width=1.9cm, minimum height=0.9cm, font=\small},
  note/.style={font=\scriptsize, align=center, fill=white, inner sep=1pt},
  arr/.style={-{Stealth[length=2.5mm]}, very thick},
]
\node[box] (efit) at (0, 2.2) {FAIR-MAST\\ $\mathbf{x}_{\text{real}}$};
\node[choice] (lao) at (3.0, 2.2) {Lao85 fit\\$p'(\psi),\,FF'(\psi)$};
\node[box, minimum width=2.8cm] (theta) at (6.2, 2.2) {Source params $\tilde{\theta}_{\text{pert}}$\\{\scriptsize sampling}};
\node[box] (gs) at (9.4, 2.2) {GS forward\\$\rightarrow\,\psi_{\text{sim}}$};
\node[box] (xreal) at (3.0, 0.8) {EFIT reconstruction\\$\psi_{\text{real}}$};
\node[choice, minimum width=2.8cm] (diag) at (9.4, 0.8) {Diagnostic $+$ noisy\\{\scriptsize calibrated from EFIT}\\$\rightarrow\,\mathbf{x}_{\text{sim}}$};
\node[box] (pr) at (3.0, -0.6) {Real pairs\\$(\mathbf{x}_{\text{real}},\psi_{\text{real}})$};
\node[box] (ps) at (9.4, -0.6) {Sim pairs\\$(\mathbf{x}_{\text{sim}},\psi_{\text{sim}})$};
\node[key] (train) at (6.2, -2.0) {Training:\\sim pre-training $+$ real fine-tuning};

\draw[arr] ([xshift=0.35cm]xreal.north) -- ([xshift=0.35cm]lao.south);
\draw[arr] (lao) -- (theta);
\draw[arr] (theta) -- (gs);
\draw[arr] (gs) -- (diag);
\draw[arr] (diag) -- (ps);
\draw[arr] (efit.south) .. controls (1.0, 1.6) and (2.0, 1.4) ..
      (xreal.north);
\draw[arr] (xreal) -- (pr);
\draw[arr] (pr.south) -- (train.north west);
\draw[arr] (ps.south) -- (train.north east);
\end{tikzpicture}%
}

\caption{Sim-to-real data generation pipeline: real MAST EFIT equilibria are parameterized by Lao85 profiles, statistically perturbed into source parameters $\tilde{\theta}_{\text{pert}}$, and solved forward with FreeGSNKE to produce synthetic equilibria; synthetic diagnostics are generated by forward modeling with EFIT-calibrated noise.}
\label{fig:pipeline}
\end{figure}

Each solved equilibrium is filtered by seven hard criteria: (1) GS nonlinear residual $\le 10^{-8}$; (2) relative error between $\int J_{\phi}$ and $I_p$ below $10^{-6}$; (3) $p(s)\ge 0$; (4) $F^2(s)>0$; (5) plasma current strictly inside the limiter mask; (6) $\overline{\psi}\in[0,1]$ within the mask; (7) limited/diverted topology consistent with the EFIT reference at the same target time (topology anchoring). These ensure physical solutions. The whole pipeline is shown in \Cref{fig:pipeline}.

\section{Reconstruction model and evaluation protocol}
\label{sec:method}

\subsection{Problem formulation}
We formulate free-boundary equilibrium reconstruction as a supervised
regression problem. Each equilibrium is reconstructed independently from
the diagnostics at a single static time slice. For each sample, we denote
the model input by $\mathbf{X}\in\mathbb{R}^{69}$ and the target by the
poloidal-flux field $\psi(R,Z)$. Specifically,

\begin{equation}
\mathbf{X}
=
\left[t, \,
I_p(t),\,
\boldsymbol{\Phi}_{\mathrm{loop}}(t),\,
\mathbf{B}_{\mathrm{pickup}}(t),\,
\mathbf{I}_{\mathrm{PF}}(t)
\right]
\longrightarrow
\hat{\psi}_{\theta}(R,Z).
\label{eq:mapping}
\end{equation}
Here, $t$ is the recorded time within each discharge,
$I_p$ is the plasma current,
$\mathbf{I}_{\mathrm{PF}}$ denotes the active poloidal-field coil
currents, and $\boldsymbol{\Phi}_{\mathrm{loop}}$ and
$\mathbf{B}_{\mathrm{pickup}}$ denote the flux-loop and magnetic-probe
measurements, respectively.

Under the quasi-static approximation, a discharge is represented
by a sequence of instantaneous equilibria~\citep{blum2012reconstruction}.
Restricting samples to the current flat-top window does not
require the poloidal-flux field to be time independent. We include the time within each discharge, $t$, as an auxiliary contextual input to allow the model to use potential statistical
associations between time and the poloidal-flux field within
the retained window.

\begin{table}[htbp]
\centering
\small
\begin{tabular}{lc}
\toprule
Input block & No. of features \\
\midrule
Discharge time $t$ & 1 \\
Plasma current $I_p$ & 1 \\
Flux-loop measurements & 10 \\
Pickup-probe signals & 46 \\
Active-coil currents & 11 \\
\midrule
Total & 69 \\
\bottomrule
\end{tabular}
\caption{Input feature vector for equilibrium reconstruction:
68 diagnostic and current features together with discharge time,
giving 69 features in total.}
\label{tab:inputs68}
\end{table}

\subsection{Reconstruction architecture}
For the reconstruction backbone, we adopt TokaMind~\citep{amorisco2026tokamind}, a multi-modal transformer foundation model for tokamak plasma dynamics pretrained on heterogeneous diagnostics from the public FAIR-MAST dataset~\citep{jackson2024fairmast}. TokaMind tokenizes input signals into embeddings, processes them with a masked Transformer encoder, and predicts outputs with task-specific heads; explicit token-availability masks make it robust to missing diagnostics. 
Note that we only adopt the TokaMind architecture instead of its pretrained checkpoint parameters. 
Instead, our models are initialized according to the training strategy
described in Section~4.3, i.e., either from random initialization or from
synthetic-data pre-training, so that we can investigate the effect of synthetic pre-training.
In this work, the model input consists of 68 diagnostic and current
features together with discharge time, giving 69 input features in
total, as summarized in Table~\ref{tab:inputs68}. These features are
concatenated into a single fusion-state input for equilibrium
reconstruction.

\begin{figure}[t]
    \centering
    \includegraphics[width=\linewidth]{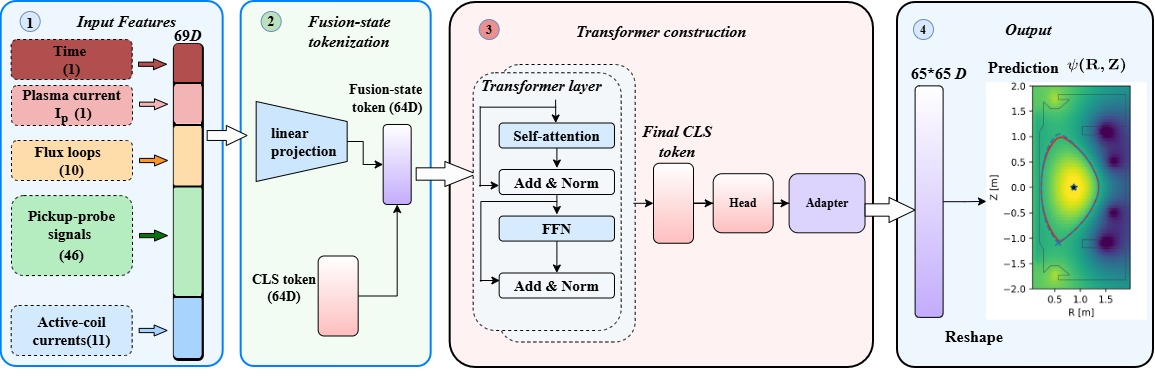}
    \caption{Overview of the TokaMind-based equilibrium reconstruction
architecture. The 68 diagnostic and current features together with
discharge time form a 69-dimensional input vector, which is embedded
into a fusion-state token and processed with a CLS token through the
Transformer backbone. The output adapter reconstructs the
$65\times65$ poloidal-flux field.}
    \label{fig:tokamind_architecture}
\end{figure}

For equilibrium reconstruction, we instantiate the TokaMind backbone with an embedding dimension of $d_{\mathrm{model}}=64$, 2 Transformer layers, 4 attention heads, a feed-forward dimension of 128, and a dropout rate of 0.05, resulting in 363,393 trainable parameters. The overall architecture of the proposed equilibrium reconstruction model is illustrated in \Cref{fig:tokamind_architecture}. 

At each static time slice, the model input is the input vector
$\mathbf{X}$ in Equation~\eqref{eq:mapping}, i.e. the 69 input features described
in \Cref{tab:inputs68} concatenated in a fixed feature order into a single
fusion-state input chunk: the 46 probes comprise 29 CCBV, 11 OBR, and 6 OBV
signals. The 69-dimensional input vector is first projected into a
$d_{\mathrm{model}}$-dimensional embedding, which constitutes a single
fusion-state input token. A learnable classification (CLS) token is then
prepended to the fusion-state token, and its final representation aggregates
the information from the input. After Transformer encoding, the CLS
representation is passed through a modality-specific prediction head followed
by an output adapter, which maps the latent representation to a
4,225-dimensional vector corresponding to the flattened $65\times65$
poloidal-flux field $\psi_{\mathrm{pred}}(R,Z)$.

\subsection{Training objective and procedure}

Given a training pair $(\mathbf{X}_{\mathrm{train}},\psi_{\mathrm{target}})$,
we minimize the mean squared error (MSE) between the predicted and target
poloidal-flux fields. All input features, including discharge time, and the target flux
fields are standardized using statistics computed exclusively
from the corresponding training set.

\begin{equation}
\mathcal{L}(\theta)
=
\frac{1}{|\mathcal{D}|}
\sum_{(\mathbf{X}_{\mathrm{train}},\psi_{\mathrm{target}})\in\mathcal{D}}
\left\|
\hat{\psi}_{\theta}(\mathbf{X}_{\mathrm{train}})
-
\psi_{\mathrm{target}}
\right\|_2^2 ,
\label{eq:training_loss}
\end{equation}
where $\hat{\psi}_{\theta}$ denotes the model prediction parameterized by
$\theta$, and $\mathcal{D}$ denotes the training set. The loss is
evaluated over the complete $65\times65$ computational grid without
additional plasma-region masking or auxiliary physics-based regularization.
During evaluation, the predicted flux fields are transformed back to the
original target scale before calculating the reconstruction metrics.

The reconstruction model is first pre-trained on the synthetic dataset
generated in \Cref{sec:gen} and then fine-tuned on real MAST data, with all
parameters remaining trainable. Each stage uses its corresponding
held-out validation set for early stopping and checkpoint selection.
Detailed optimization settings are provided in Appendix~\ref{app:training_details}.

\section{Experimental design}
\label{sec:experimental_design}

\subsection{Comparison and data budgets controls}
Considering the scarcity and cost of experimentally reconstructed labels, we mainly investigate whether inexpensive simulations can provide a physical prior that improves learning from limited real data.
At a real-data fraction $r$, both arms use the same $rN$ real training samples,
where $N$ is the size of the full real training partition, while the
pre-trained arm additionally uses $N_{\mathrm{syn}}$ synthetic samples whose
parent discharges belong to that partition. Both training strategies use the same architecture and the same
input definition, including discharge time. The comparisons
therefore evaluate the quantitative benefit of synthetic pre-training for
a fixed reconstruction task and input specification.

\begin{table}[htbp]
\centering
\small
\begin{tabular}{lccc}
\toprule
Training strategy & Synthetic pre-training & Real training & Total \\
\midrule
Real from scratch & $0$ & $rN$ & $rN$ \\
Pre-train $+$ fine-tune & $N_{\mathrm{syn}}$ & $rN$ & $N_{\mathrm{syn}}+rN$ \\
\bottomrule
\end{tabular}
\caption{Data-budget accounting for our method and the baseline. At each real-data fraction $r$, both strategies use the same $rN$ real samples. The pre-trained strategy additionally uses $N_{\mathrm{syn}}$ synthetic samples generated from parent discharges in the training partition.}
\label{tab:data_budget_controls}
\end{table}

\subsection{Generalization settings}
To assess the benefit of synthetic pre-training, we use as baseline a model trained from scratch on real data, and compare it with a model initialized from synthetic pre-training and then fine-tuned on the same real data; all gains reported below are measured relative to this from-scratch baseline. In every setting both arms are evaluated at real-data fractions of $1/5/25/50/100\%$ obtained by shot-level subsampling.
Training, validation, and test partitions are mutually disjoint
at the discharge level, and pre-training uses only synthetic
instances whose parent discharges belong to the training partition,
so that no validation- or test-parent equilibrium is supplied
to the reconstruction model during pre-training. The three settings differ in the type of generalization they demand (shot/slice counts are given in \Cref{tab:dataset_splits}):
\begin{itemize}
\item \textbf{Random split.} The standard, independent and identically distributed baseline: discharges are partitioned at random into train/validation/test. Because train and test come from the same distribution, this setting isolates the plain in-distribution behaviour of pre-training, where the advantage is expected to be more significant as real data become scarcer.

\item \textbf{Shape-OOD.} The identical-distribution assumption is dropped: the test set is enriched in plasma shapes that are rare in, or absent from, the training set, defined from the real equilibria alone via their upper--lower triangularity asymmetry. The Shape-OOD protocol is constructed only from diverted equilibria with valid upper- and lower-triangularity descriptors; samples that do not satisfy these requirements are excluded before the discharge-level split. This probes whether the pre-trained prior generalises to out-of-distribution plasma shapes rather than merely fitting the training manifold.

\item \textbf{Temporal-OOD.} The application-oriented, most demanding setting: discharges are partitioned by campaign, so that the model is pre-trained and fine-tuned on past campaigns (M5--M7) and must reconstruct discharges of a future, unseen campaign (M9). This mirrors the practical scenario of deploying a surrogate at the start of a new experimental campaign, before any of its equilibrium labels exist.
\end{itemize}

\begin{table}[htbp]
\centering
\small
\begin{tabular}{lccc}
\toprule
Split & Train (shots/slices) & Val (shots/slices) & Test (shots/slices) \\
\midrule
Random    & 5{,}828 / 171{,}427 & 728 / 21{,}147 & 730 / 21{,}350 \\
Shape-OOD & 5{,}743 / 176{,}319 & 170 / 2{,}835 & 775 / 19{,}009 \\
Temporal  & 6{,}046 / 187{,}292 & 559 / 6{,}206 & 507 / 17{,}498 \\
\bottomrule
\end{tabular}
\caption{Number of discharges and time slices in each of the three evaluation splits.}
\label{tab:dataset_splits}
\end{table}

\subsection{Evaluation metrics}
We evaluate reconstruction quality at two levels: at the data level, we employ standard machine-learning metrics including the root mean squared error (RMSE), the per-sample normalized RMSE (nRMSE), and the normalized mean absolute error (RMAE) on the predicted poloidal flux; at the physics level, we focus on derived quantities: the toroidal current density $J_{\phi}$ recovered from the Grad--Shafranov equation and the last-closed-flux surface (LCFS) extracted from the predicted flux field. These tasks are covered in the Group~1 tasks of flux reconstruction, current-profile estimation, and boundary detection in TokaMark~\citep{rousseau2026tokamark}. Results are averaged over three independent random seeds.

\section{Results}
\label{sec:results}

We present results organised by the three evaluation settings of \Cref{sec:experimental_design}: the in-distribution \emph{random split}, the \emph{shape out-of-distribution} (Shape-OOD) split, and the \emph{cross-campaign} temporal-OOD.

\subsection{Random split: sim-to-real gap and fine-tuning}
\begin{figure}
    \centering
    \includegraphics[width=0.5\linewidth]{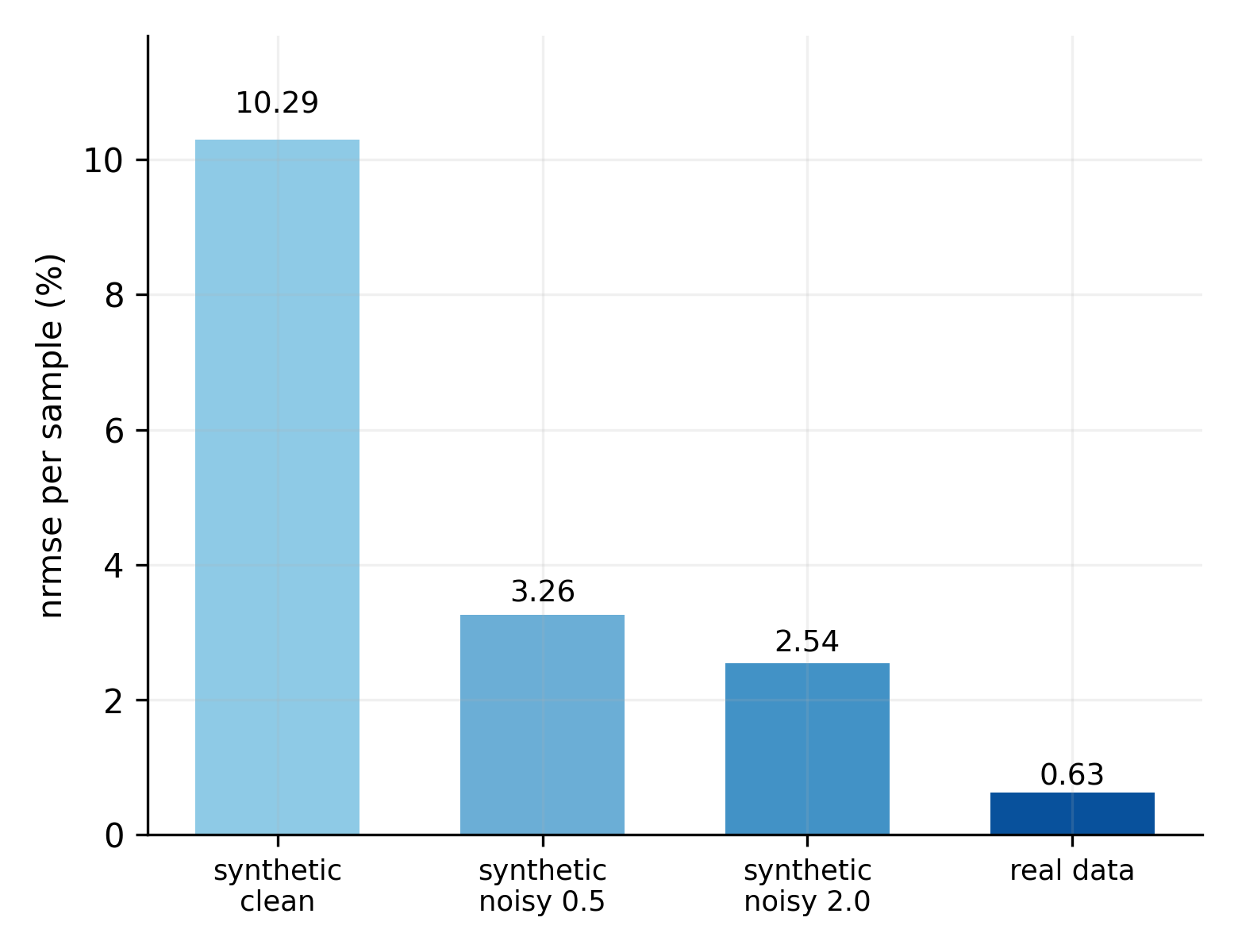}
    \caption{Zero-shot evaluation of synthetic-only models trained under clean and noisily corrupted variants on real discharges reveals a pronounced sim-to-real gap.}
    \label{fig:sim2real_gap}
\end{figure}

We first report the random split results. A prominent sim‑to‑real gap is demonstrated by zero‑shot evaluation as shown in \Cref{fig:sim2real_gap}, where the model trained on clean synthetic data yields the highest nRMSE when directly tested on real discharges. Injecting simulated diagnostic noise into synthetic training data partially mitigates this domain mismatch. Nevertheless, all synthetic‑only models still underperform the model trained on all real data, confirming that synthetic data alone cannot fully reproduce real‑world measurement characteristics. 

After fine-tuning on real experimental data, the benefit of synthetic pre-training is clearly observed in the random-split in-distribution setting, as shown in \Cref{tab:random_table}. The gain grows as real data become scarcer, reaching roughly $+10$--$12\%$ at 5\% and $+22$--$29\%$ at 1\% real data across the three noise variants, and it fades as the amount of real data increases, becoming negligible or slightly negative at fractions of 25\% and above. 
The benefit is largely insensitive to the injected diagnostic noise: the clean and noisily pre-trained arms yield nearly identical fine-tuned accuracy, with nRMSE differing by less than $0.1$ at scarce fractions.

\begin{table}[htbp]
\centering
\small
{\setlength{\tabcolsep}{3pt}
\begin{tabular}{lccccccc}
\toprule
\shortstack{Real-data\\fraction} & Real scratch & \multicolumn{2}{c}{Clean} & \multicolumn{2}{c}{Noisy 0.5} & \multicolumn{2}{c}{Noisy 2} \\
\cmidrule(lr){2-2} \cmidrule(lr){3-4} \cmidrule(lr){5-6} \cmidrule(lr){7-8}
& nRMSE & nRMSE & Gain & nRMSE & Gain & nRMSE & Gain \\
\midrule
1\%   & $1.483 \pm 0.039$ & $1.076 \pm 0.010$ & $\bm{+27.5\%}$ & $1.057 \pm 0.016$ & $\bm{+28.7\%}$ & $1.147 \pm 0.038$ & $\bm{+22.6\%}$ \\
5\%   & $1.107 \pm 0.022$ & $0.974 \pm 0.012$ & $\bm{+12.0\%}$ & $0.981 \pm 0.022$ & $\bm{+11.4\%}$ & $0.991 \pm 0.011$ & $\bm{+10.5\%}$ \\
25\%  & $0.772 \pm 0.039$ & $0.815 \pm 0.021$ & $-5.6\%$ & $0.809 \pm 0.040$ & $-4.8\%$ & $0.807 \pm 0.024$ & $-4.6\%$ \\
50\%  & $0.714 \pm 0.013$ & $0.729 \pm 0.013$ & $-2.2\%$ & $0.764 \pm 0.037$ & $-7.0\%$ & $0.749 \pm 0.030$ & $-4.9\%$ \\
100\% & $0.659 \pm 0.025$ & $0.686 \pm 0.032$ & $-4.1\%$ & $0.658 \pm 0.011$ & \textcolor{gray!60}{$+0.1\%$} & $0.653 \pm 0.003$ & \textcolor{gray!60}{$+0.9\%$} \\
\bottomrule
\end{tabular}}
\caption{Per-sample nRMSE (\%) on the random-split test set across five real-data fractions, comparing from-scratch training on real data (Real scratch) with synthetic pre-training plus real fine-tuning under three noise conditions (Clean, Noisy 0.5, Noisy 2). Gain is the relative improvement over the from-scratch baseline, Gain $=(E_{\mathrm{scratch}} - E_{\mathrm{pretrained}})/E_{\mathrm{scratch}}\times100\%$. Results are mean $\pm$ std over three seeds. Bold gains indicate a statistically significant improvement from pre-training; gains shown in gray are positive but negligible.}
 \label{tab:random_table}
\end{table}

\begin{figure*}[t]
    \centering
    \includegraphics[width=0.92\textwidth]{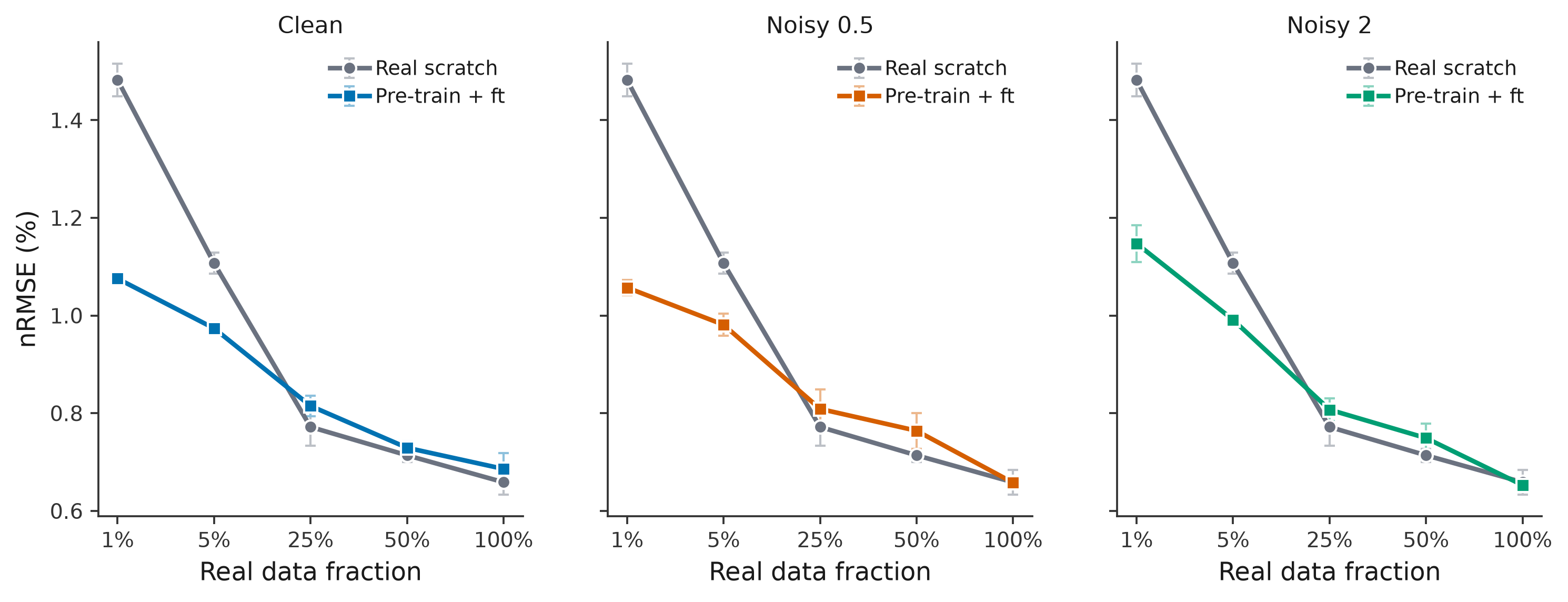}
    \caption{Per-sample nRMSE (\%) on the random-split test set versus the real-data fraction, comparing from-scratch training on real data (Real scratch) with synthetic pre-training plus real fine-tuning under three pre-training noise conditions (Clean, Noisy 0.5, Noisy 2). The gap between the two curves is the sim-to-real gain; it is confined to the scarce-data regime and is no longer beneficial once abundant in-domain labels are available. Lower is better.}
    \label{fig:random_split_curves}
\end{figure*}

\begin{figure}[htbp]
\centering

\begin{subfigure}{\textwidth}
\centering
\includegraphics[width=0.85\textwidth]{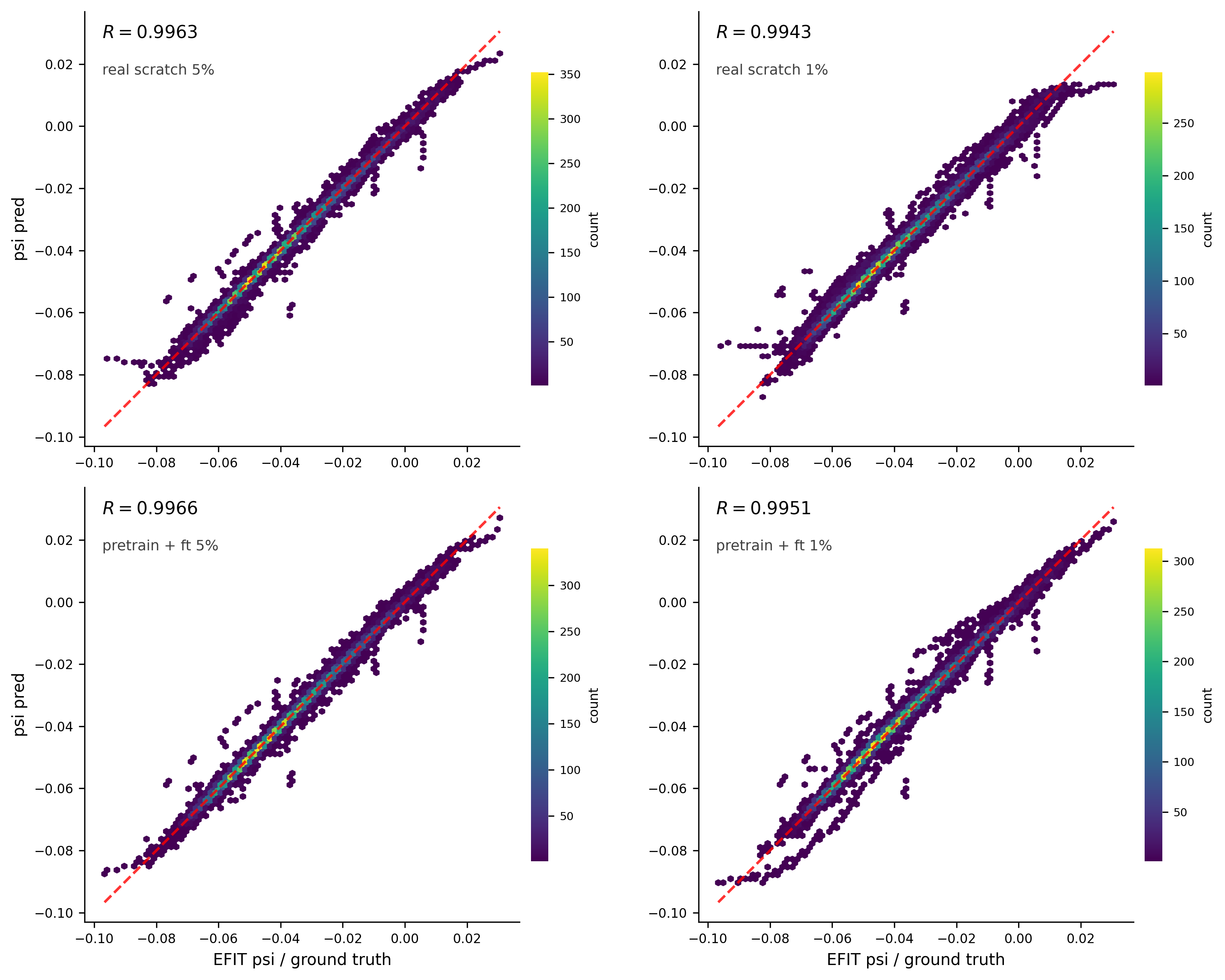}
\caption{For each panel, the per-sample mean $\psi$ averaged over the $65\times 65$ grid is shown for predicted vs EFIT values across the test set of $21{,}350$ points per panel. Left column: 5\% data fraction. Right column: 1\% data fraction.}
\label{fig:psi_scatter_mean}
\end{subfigure}
\vspace{0.3cm}

\begin{subfigure}{\textwidth}
\centering
\includegraphics[width=0.85\textwidth]{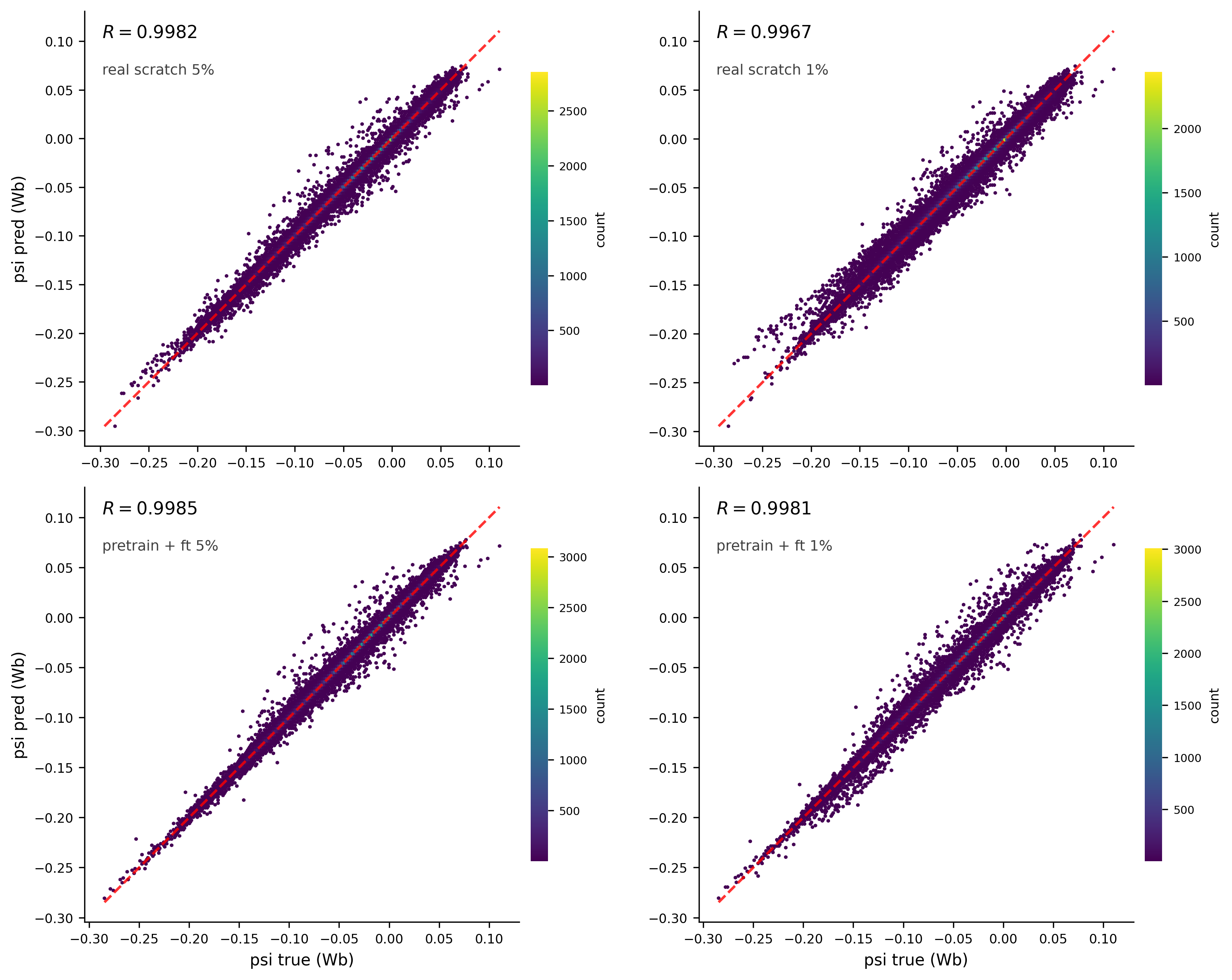}
 \caption{Because the $65\times65$ grid across the full test set yields a very large number of points, about $21{,}350\times65^2\approx9\times10^7$, we visualize only a uniform sample: $\psi$ values are drawn at grid points with probability $0.2\%$ and a fixed seed, giving about $1.8\times10^5$ points per panel.}
\label{fig:psi_scatter_point}
\end{subfigure}

\caption{Predicted vs. EFIT $\psi$ scatter for the real data baseline and pretrained + fine-tuned model at 5\% and 1\% data fractions. Left column: 5\% data fraction. Right column: 1\% data fraction.}
\label{fig:psi_scatter}
\end{figure}

\Cref{fig:psi_scatter} confirms that both models track the equilibrium flux well, and that pretrained model tightens agreement with the ground truth at the individual grid-point level, most notably in the 1\% regime. 

\begin{figure}
    \centering
    \includegraphics[width=0.85\linewidth]{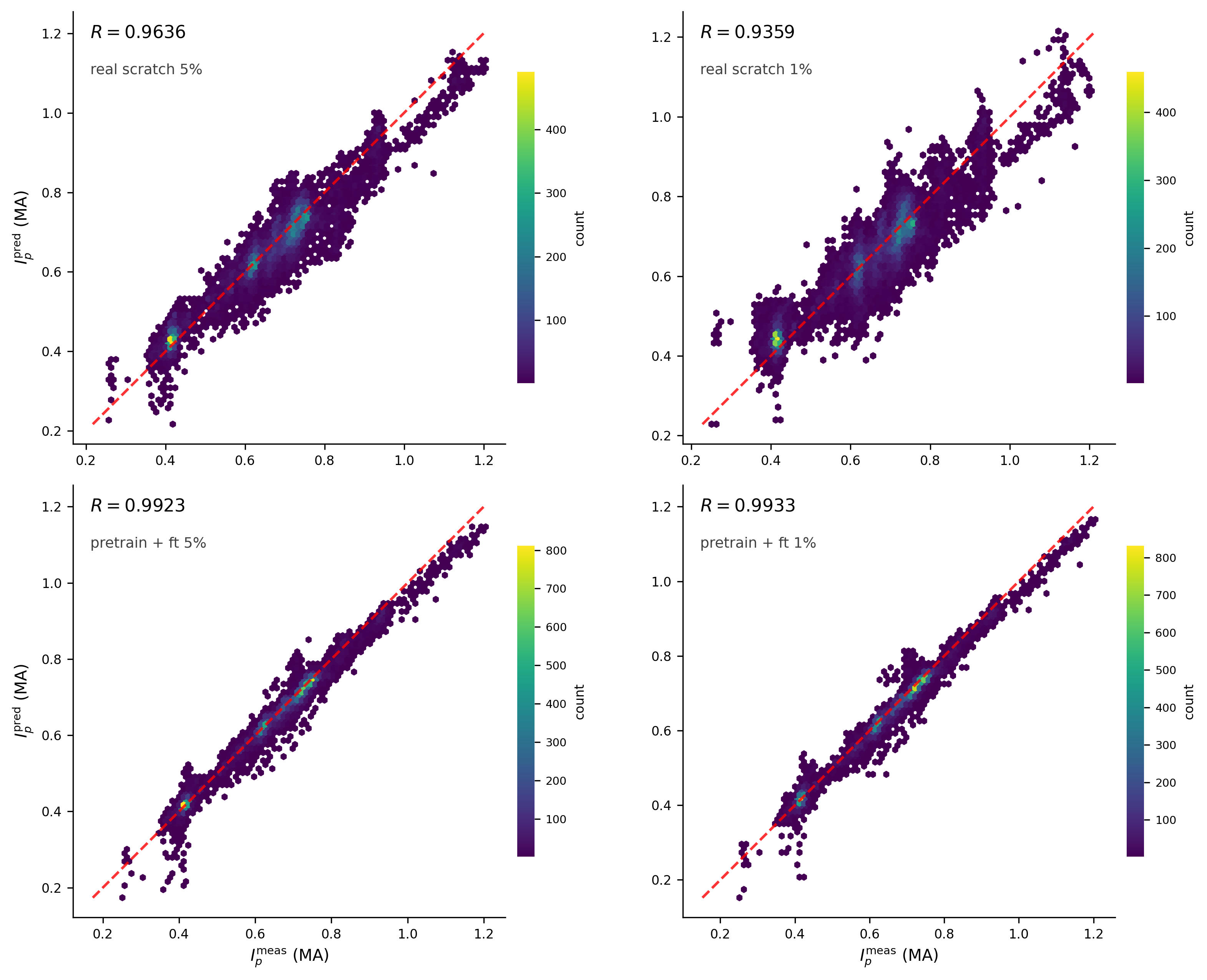}
    \caption{Predicted vs.\ measured plasma current $I_p$ per test time slice ($21{,}350$ points per panel) for the from-scratch and pre-trained + fine-tuned models at 5\% and 1\% data fractions.}\label{fig:jtor_ip_combined_1_5pct}
\end{figure}

To verify the physical consistency of the reconstructed equilibrium, we evaluate the surface integral of $J_{\phi} = -\Delta^{*}\psi/(\mu_0R)$ over the EFIT-defined plasma region to get predicted plasma current $I_p$. The predicted $I_p$ is then compared against the directly measured Rogowski-coil signal. \Cref{fig:jtor_ip_combined_1_5pct} shows the physical consistency. 

\begin{figure}[htbp]
\centering
\includegraphics[width=0.9\textwidth]{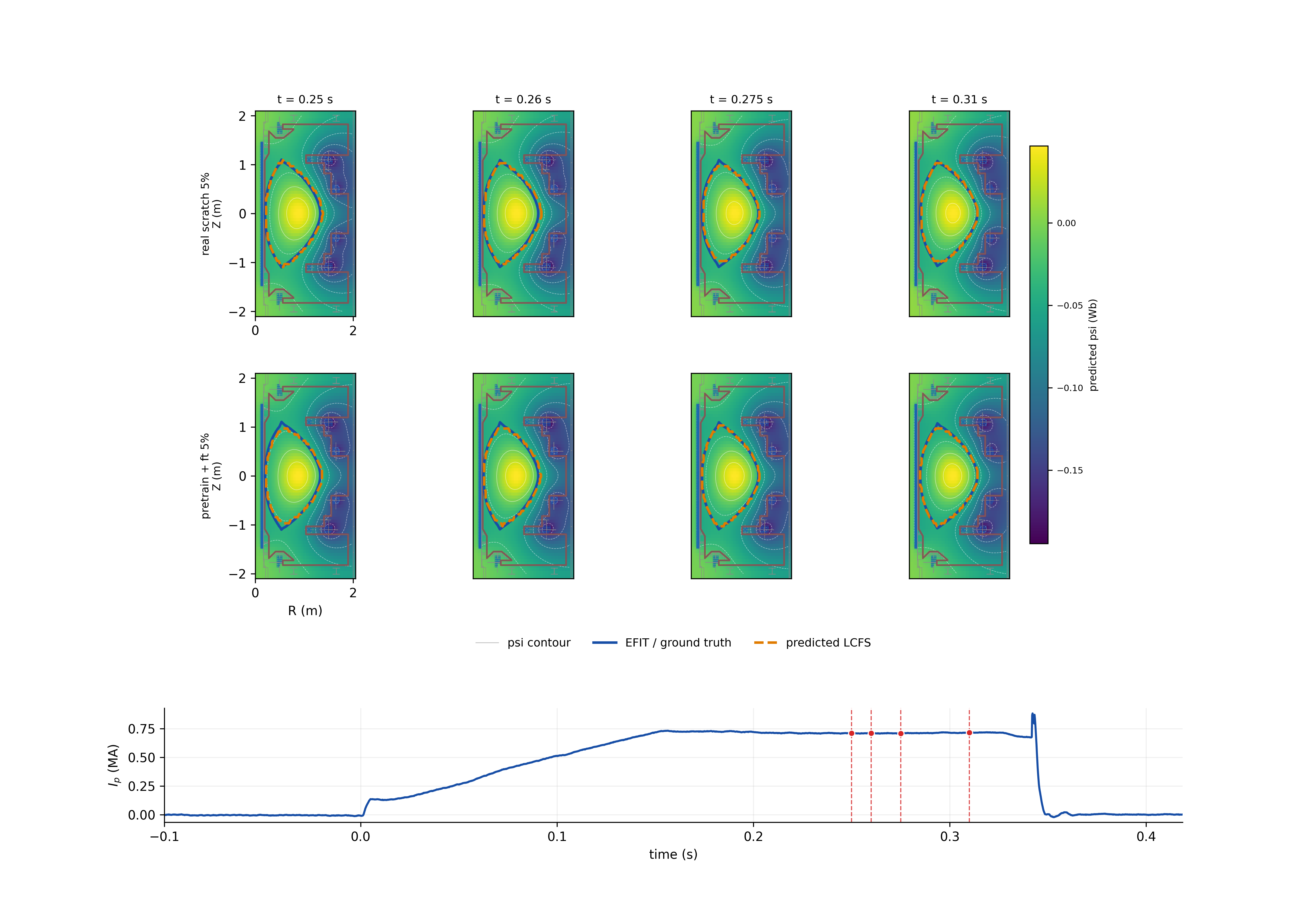}
\caption{LCFS comparison for shot 17833 at four time slices, using a model fine-tuned on 5\% real data. Top: predicted $\psi$ contours with EFIT LCFS (solid blue) and predicted LCFS (dashed orange). Bottom: $I_p$ waveform with selected slices marked.}
\label{fig:lcfs_17833_5pct}
\end{figure}

\begin{figure}[htbp]
\centering
\includegraphics[width=0.9\textwidth]{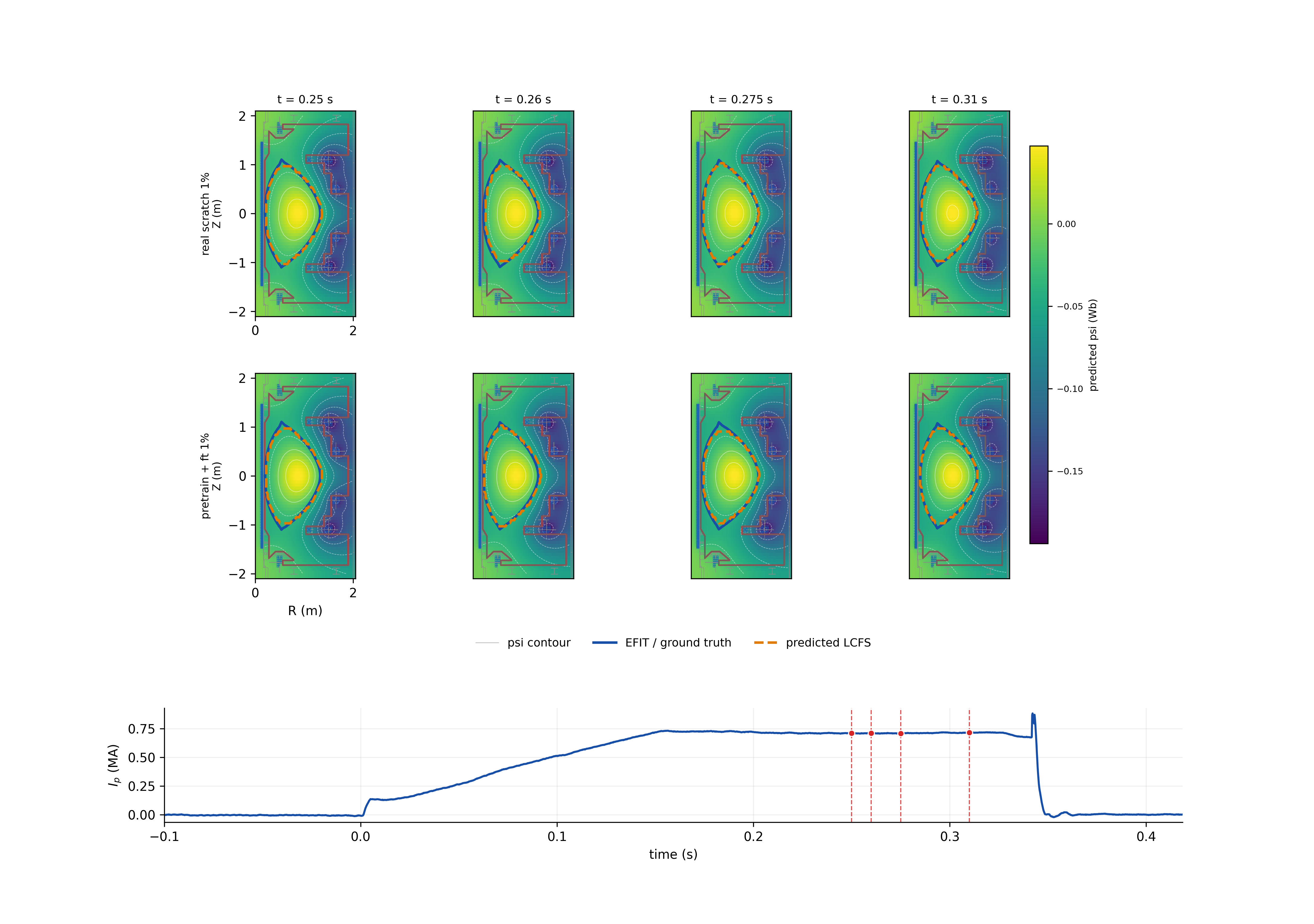}
\caption{LCFS comparison for shot 17833 at four time slices, using a model fine-tuned on 1\% real data.}
\label{fig:lcfs_17833_1pct}
\end{figure}

\Cref{fig:lcfs_17833_5pct,fig:lcfs_17833_1pct} compare the LCFS predictions of the from-scratch and pre-trained models on shot 17833 at 5\% and 1\% real data, respectively. While the from-scratch model shows increasing boundary distortion as data decrease from 5\% to 1\%, the pre-trained model maintains accurate LCFS predictions across both fractions, confirming that synthetic pre-training preserves the plasma boundary geometry even under extreme data scarcity.

These results show that the benefit of pre-training appears only where real data are scarce and turns slightly negative once abundant in-domain labels are available, where a fully converged from-scratch model is already saturated. This is expected when the test distribution matches a well-labeled training set, and it highlights that a transferable prior pays off precisely under label scarcity. In the following subsections, we turn to settings with both distribution shift and label scarcity, where pre-training provides more significant gains at every data fraction.

\subsection{Shape-OOD: out-of-distribution generalization}

To evaluate reconstruction performance for plasma shapes that are poorly
represented by the experimental training set, a dedicated shape
out-of-distribution (Shape-OOD) protocol is constructed from geometric
descriptors of the last closed flux surface (LCFS). Shape extrapolation has previously been used to assess Grad--Shafranov surrogates by extending equilibrium shape parameters beyond their training ranges~\cite{ding2026physicsinformedicml}. Experimental equilibria can exhibit appreciable up--down asymmetry in the LCFS geometry. Accordingly, following the upper and lower triangularity descriptors used for experimentally reconstructed plasma boundaries~\cite{stokolesov2025neural}, we define:
\begin{equation}
\delta_u=
\frac{R_{\rm geo}-R_{\rm upper}}{a},
\qquad
\delta_l=
\frac{R_{\rm geo}-R_{\rm lower}}{a},
\label{eq:upper_lower_triangularity}
\end{equation}
where $R_{\rm geo}=(R_{\max}+R_{\min})/2$ and
$a=(R_{\max}-R_{\min})/2$ denote the geometric major radius and horizontal
minor radius of the LCFS, respectively, while $R_{\rm upper}$ and
$R_{\rm lower}$ are the major-radius coordinates of its uppermost and
lowermost points. The upper and lower triangularity asymmetry is expressed as:
\begin{equation}
\delta_{\rm asym}=\delta_u-\delta_l.
\label{eq:delta_asym}
\end{equation}
Thus, $\delta_{\rm asym}\simeq0$ indicates comparable upper and lower triangularities, whereas increasingly negative values indicate stronger lower-triangularity dominance. These shape quantities are used
only for constructing and analysing the OOD benchmark and are not
provided to the reconstruction model as input features.

For the diverted-plasma Shape-OOD experiment, the OOD region is defined solely from the empirical distribution of the real data as:
\begin{equation}
\delta_{\rm asym}<P_{10}=-0.214,
\label{eq:shape_ood_threshold}
\end{equation}
where $P_{10}$ denotes the 10th percentile of the real diverted
equilibrium distribution. Since neighbouring equilibrium slices from the
same discharge are strongly correlated, the data partition is performed
at the discharge level rather than at the individual time-slice level.
For each discharge $s$, the OOD occupancy is defined as:
\begin{equation}
r_{\rm OOD}(s)
=
\frac{N_{\rm OOD}(s)}{N_{\rm valid}(s)}.
\label{eq:r_ood}
\end{equation}
Discharges are then stratified by $r_{\rm OOD}$ such that the training set
contains few OOD configurations, whereas the test set is enriched in them. The training, validation, and test sets are mutually disjoint at the
discharge level. The resulting OOD test set contains 19,009 equilibrium slices from
775 discharges.

\begin{figure}
	\centering
	\subfloat[]{
		\includegraphics[width=70mm]{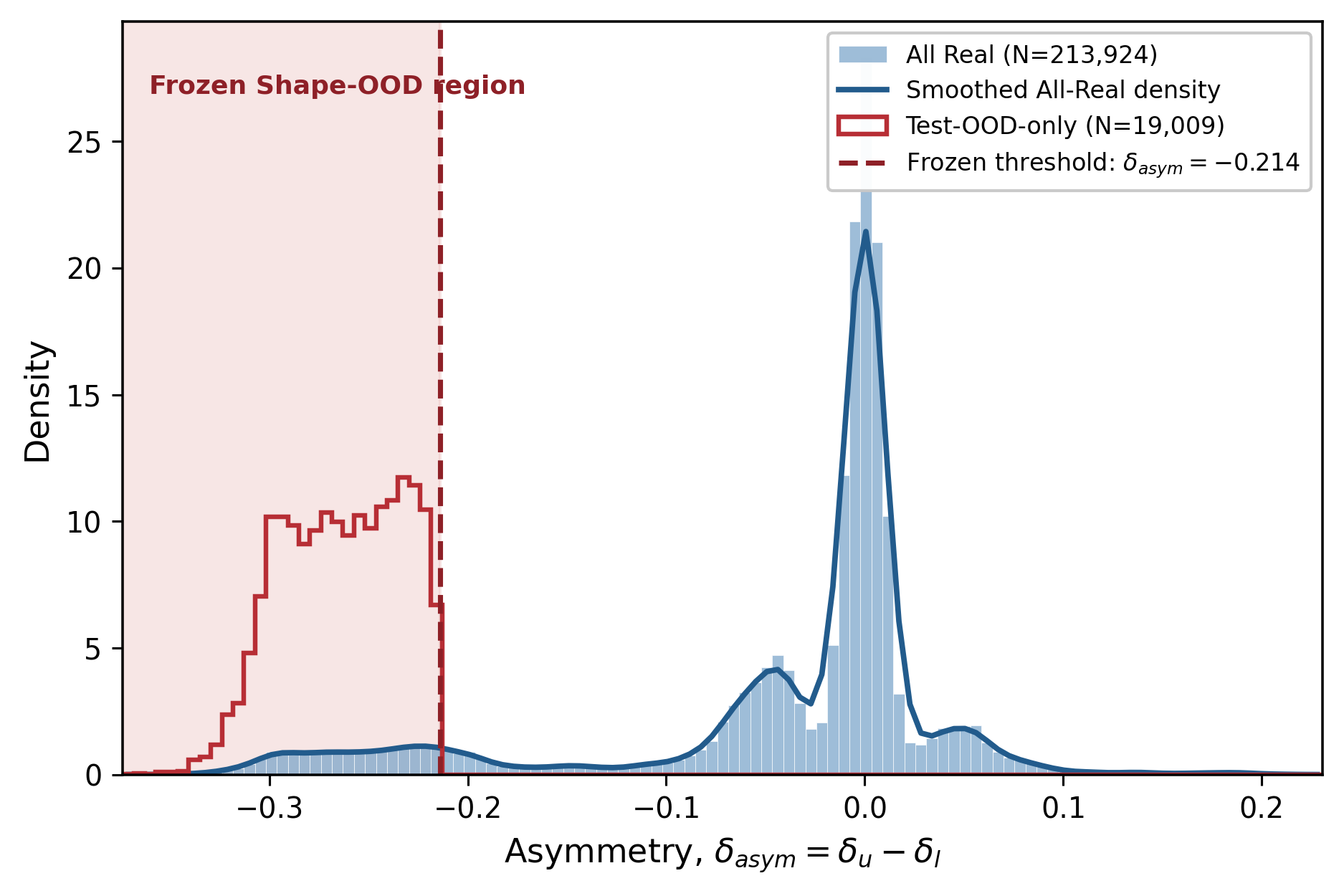}
		\label{fig:delta_asym}
	} 
	\subfloat[]{
		\includegraphics[width=70mm]{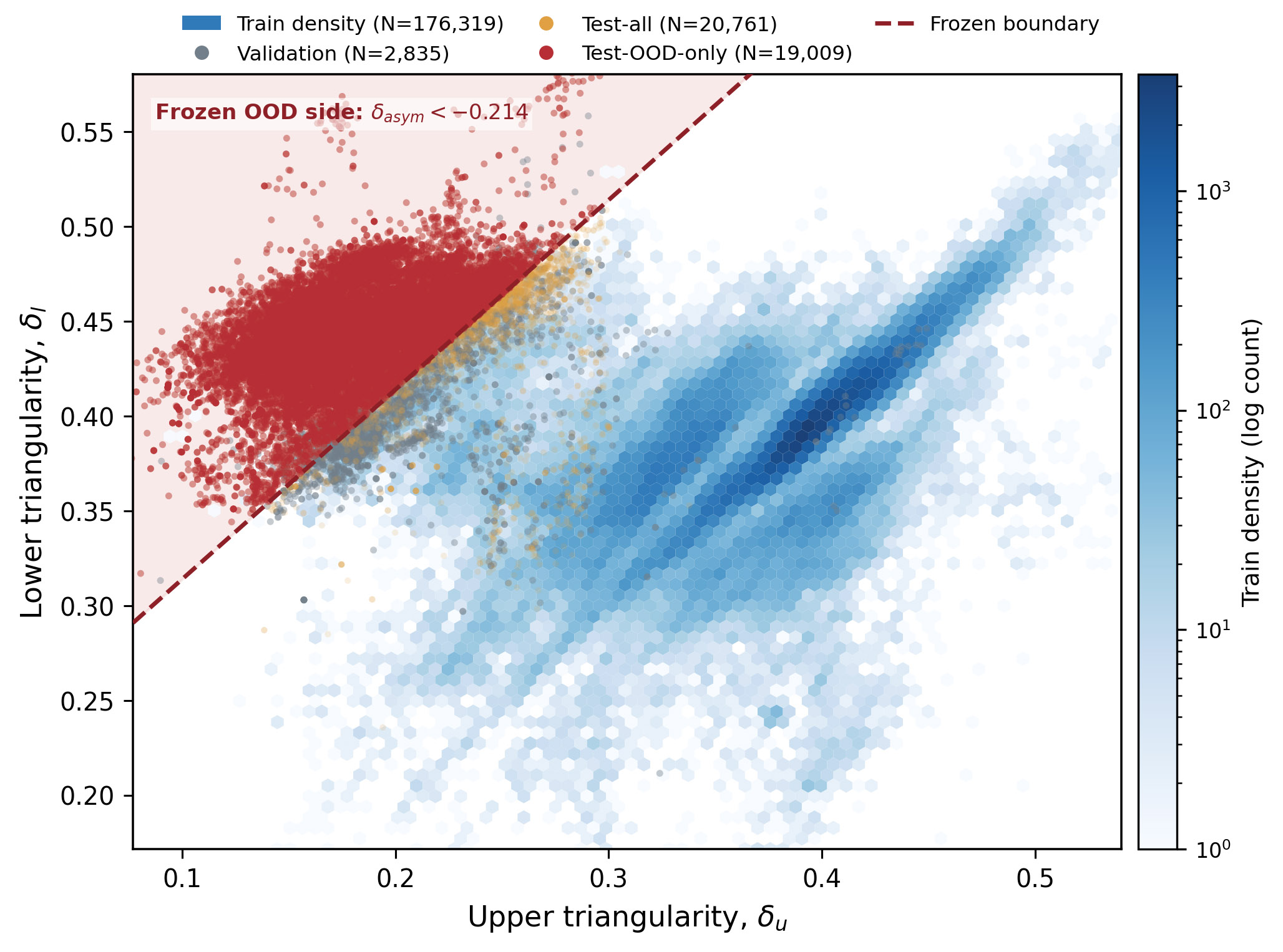}
		\label{fig:shape_split}
	}
	\caption{
	Definition of the Shape-OOD benchmark.
	(a) Distribution of the triangularity asymmetry
	$\delta_{\rm asym}=\delta_u-\delta_l$, with the frozen OOD threshold
	$\delta_{\rm asym}=-0.214$.
	(b) Training, validation, and OOD test distributions in the
$(\delta_u,\delta_l)$ shape space; the dashed line denotes the
Shape-OOD boundary.
	}
	\label{fig:shape_ood_definition}
\end{figure}

As shown in \Cref{fig:delta_asym}, the real MAST equilibria are
concentrated around $\delta_{\rm asym}=0$, indicating relatively similar
upper and lower triangularities for much of the experimental dataset.
A negative tail corresponding to stronger lower-triangularity dominance
is isolated by the threshold $\delta_{\rm asym}=-0.214$, defining the OOD
test region.
\Cref{fig:shape_split} shows the corresponding training, validation,
and test distributions in the $(\delta_u,\delta_l)$ shape space. The training samples mainly occupy the dominant experimental manifold,
whereas the OOD test samples are concentrated in the
lower-triangularity-dominant region beyond the OOD boundary.

\begin{figure}[t]
    \centering
    \includegraphics[width=70mm]
    {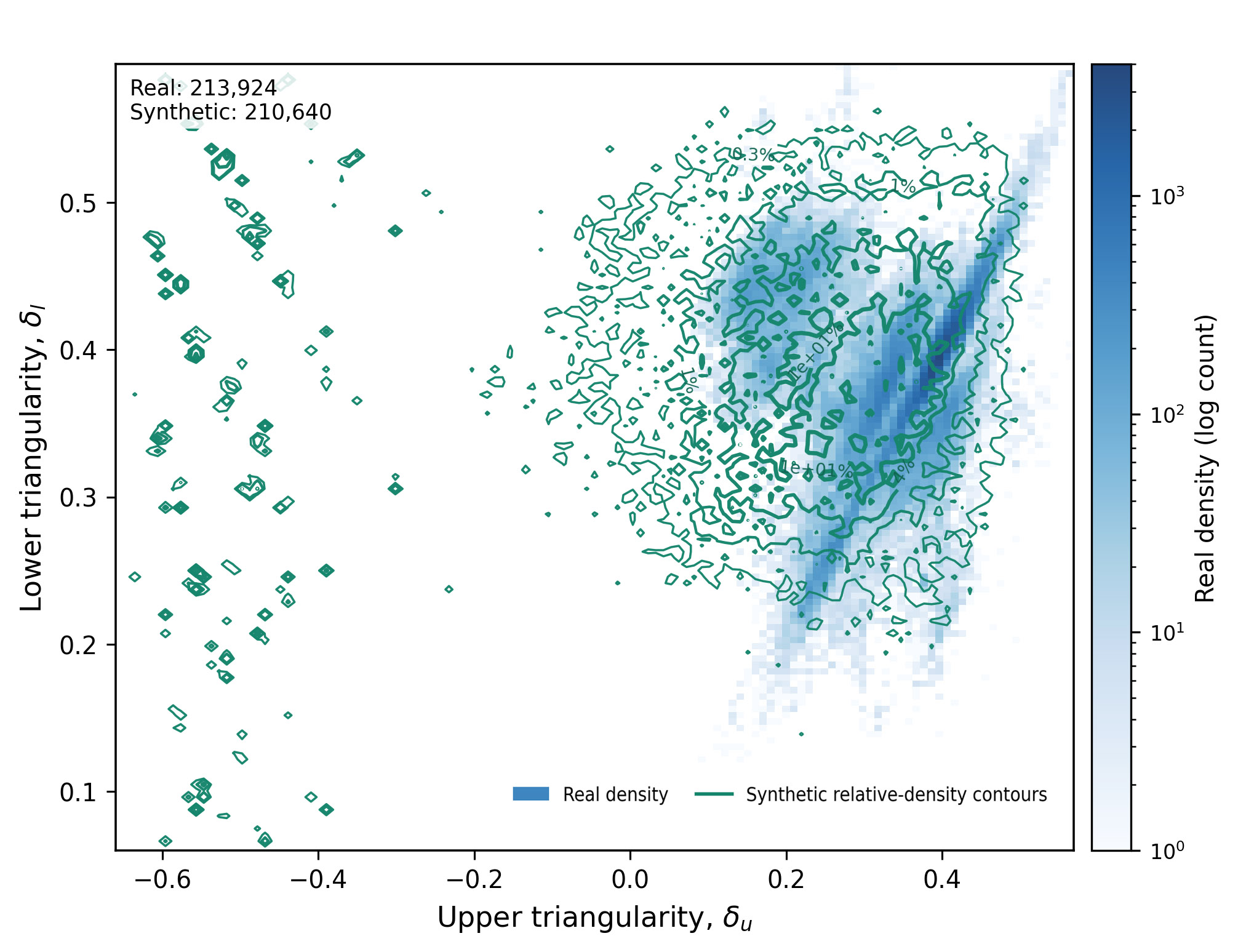}
    \caption{
    Real and synthetic equilibrium coverage in the
    $(\delta_u,\delta_l)$ shape space. The blue density represents the
    distribution of 213,924 real MAST equilibria, while the green
    contours denote the relative density of 210,640
    FreeGSNKE-generated synthetic equilibria with valid shape
    descriptors.
    }
    \label{fig:shape_ood_coverage}
\end{figure}

\Cref{fig:shape_ood_coverage} compares the real and synthetic shape
distributions. The dominant synthetic distribution substantially overlaps the
high-density region of the real equilibria, indicating that the generated
dataset captures the principal experimental shape distribution in
$(\delta_u,\delta_l)$ space. Furthermore, the synthetic distribution exhibits broader support in the $(\delta_u,\delta_l)$ space, extending into regions sparsely populated by real data.

\begin{figure*}[t]
    \centering
    \includegraphics[width=\textwidth]
    {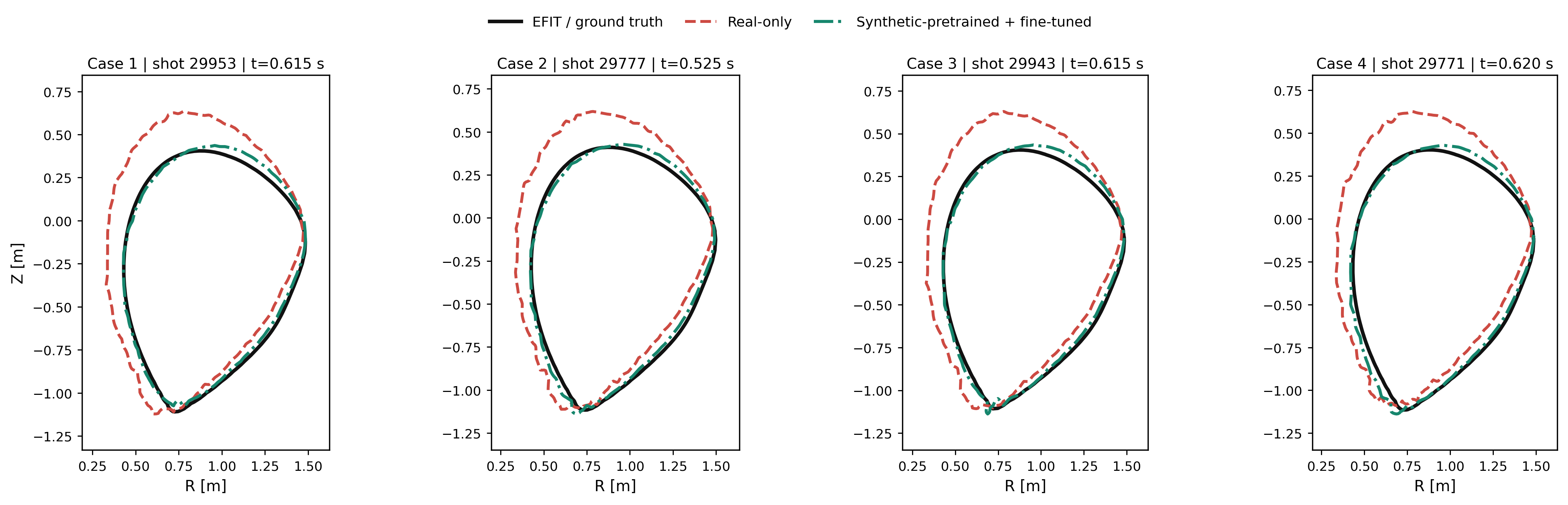}
    \caption{ Representative Shape-OOD LCFS reconstructions at the 5\% real-data
setting for four equilibria from the OOD test set, comparing the EFIT
reference, real-from-scratch reconstruction, and synthetic-pretrained
reconstruction after real-data fine-tuning.
    }
    \label{fig:shape_ood_lcfs}
\end{figure*}

This broader synthetic support provides a plausible mechanism for improved robustness under plasma-shape distribution shift.
Physics-consistent perturbations of the free-boundary equilibrium expose the reconstruction model during pre-training to a wider range of admissible plasma geometries than is available from the experimental training set alone. \Cref{fig:shape_ood_lcfs} presents four representative equilibria
selected from the OOD test set. In these representative cases, the model trained from scratch on real data exhibits noticeable LCFS distortions. The discrepancies are particularly visible near the upper plasma boundary, where the real-only predictions tend to deviate outward from the EFIT reference and produce less asymmetric boundary shapes.
In contrast, the model initialized by synthetic pre-training and
subsequently fine-tuned on real data follows the EFIT reference much more
closely over the inboard, outboard, upper, and lower boundary regions. In particular, the lower-triangularity-dominant geometry of these Shape-OOD configurations is better preserved after synthetic pre-training. The improvement therefore extends beyond global poloidal-flux accuracy to the physically relevant plasma-boundary geometry. \Cref{tab:shape_ood_results} shows that the synthetic-pretrained model achieves lower mean nRMSE and LCFS boundary error across all evaluated real-data fractions.
The improvement is most pronounced in the 1\% real-data regime and generally decreases as more experimental training data become available. Nevertheless, the pretrained model retains lower mean errors across all
evaluated fractions, indicating improved robustness under plasma-shape
distribution shift.

\begin{table*}[t]
\centering
\small
{\setlength{\tabcolsep}{5pt}
\begin{tabular}{lcccccc}
\toprule
\shortstack{Real-data\\fraction} & \multicolumn{3}{c}{nRMSE (\%)} & \multicolumn{3}{c}{LCFS distance (m)} \\
\cmidrule(lr){2-4} \cmidrule(lr){5-7}
& Real scratch & \shortstack{Pre-train +\\fine-tune} & Gain & Real scratch & \shortstack{Pre-train +\\fine-tune} & Gain \\
\midrule
1\%   & $5.383 \pm 0.019$ & $\bm{2.285 \pm 0.202}$ & $\bm{+57.6\%}$ & $0.1372 \pm 0.0008$ & $\bm{0.0317 \pm 0.0035}$ & $\bm{+76.9\%}$ \\
5\%   & $2.447 \pm 0.126$ & $\bm{1.706 \pm 0.142}$ & $\bm{+30.3\%}$ & $0.0428 \pm 0.0030$ & $\bm{0.0250 \pm 0.0014}$ & $\bm{+41.6\%}$ \\
25\%  & $2.071 \pm 0.094$ & $\bm{1.721 \pm 0.107}$ & $\bm{+16.9\%}$ & $0.0359 \pm 0.0010$ & $\bm{0.0271 \pm 0.0014}$ & $\bm{+24.6\%}$ \\
50\%  & $2.033 \pm 0.098$ & $\bm{1.674 \pm 0.206}$ & $\bm{+17.7\%}$ & $0.0332 \pm 0.0013$ & $\bm{0.0273 \pm 0.0018}$ & $\bm{+17.6\%}$ \\
100\% & $2.007 \pm 0.079$ & $\bm{1.858 \pm 0.273}$ & $\bm{+7.5\%}$ & $0.0336 \pm 0.0007$ & $\bm{0.0285 \pm 0.0028}$ & $\bm{+15.4\%}$ \\
\bottomrule
\end{tabular}}
\caption{Shape-OOD reconstruction performance on the OOD test set across five
real-data fractions, comparing training from scratch on real data with
synthetic pre-training followed by real-data fine-tuning. Gain is the relative improvement over the from-scratch baseline, Gain $=(E_{\mathrm{scratch}} - E_{\mathrm{pretrained}})/E_{\mathrm{scratch}}\times100\%$. Results are mean $\pm$ std over three seeds. Bold marks the better (lower-error) value per row, and bold gains denote improvement over the baseline.}
 \label{tab:shape_ood_results}
\end{table*}

\begin{figure}[t]
    \centering
    \includegraphics[width=0.5\columnwidth]{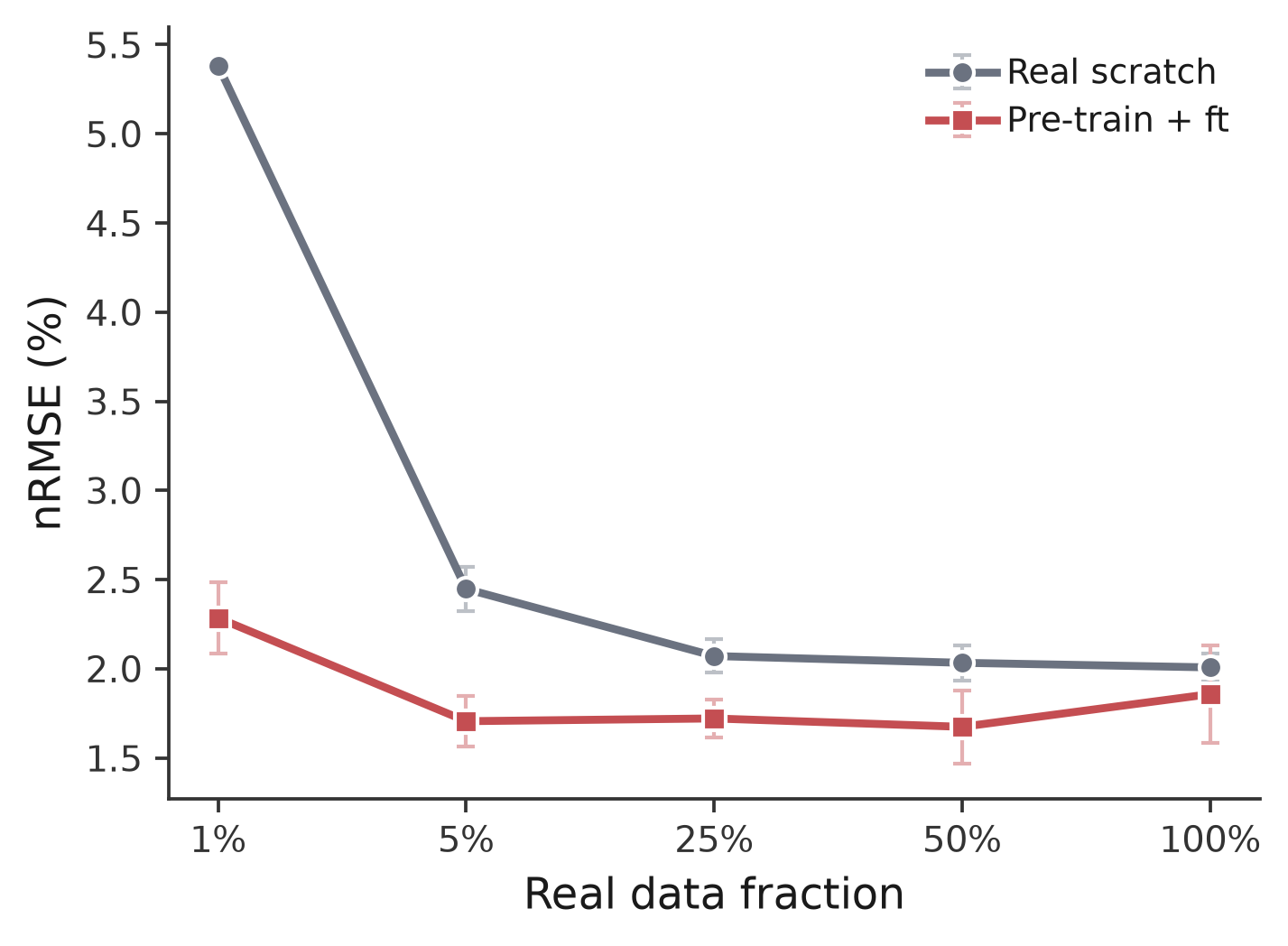}
    \caption{Per-sample nRMSE (\%) on the Shape-OOD test set versus the real-data fraction, comparing from-scratch training on real data (Real scratch) with synthetic pre-training plus real fine-tuning. The gap between the curves is the gain provided by synthetic pre-training, which is largest in the 1\% data regime and persists at every fraction. Lower is better.}
    \label{fig:shape_ood_curves}
\end{figure}

\subsection{Temporal-OOD: cross-campaign generalization}
The temporal-split protocol is motivated by a practical deployment scenario in tokamak operation: a reconstruction model trained on historical experimental campaigns must generalize to a future campaign before any labeled equilibrium data from that campaign becomes available. Cross-campaign temporal evaluation has been adopted in prior work \citep{rousseau2026tokamark} to assess generalization across successive experimental campaigns. Tokamak campaigns evolve over time due to hardware upgrades, changing operational scenarios, and diagnostic drift, so the test distribution differs from the training distribution in both plasma shape and device-level characteristics. This makes cross-campaign temporal extrapolation the most stringent generalization protocol among the three evaluated in this work. In our setup, models are trained on campaigns M5--M7 (2005--2010), validated on M8 (2011--2012), and evaluated exclusively on M9 (2013), simulating the realistic scenario of deploying a pre-trained reconstruction model at the start of a new experimental campaign when high-quality EFIT labels are not yet available.

\Cref{tab:temporal_table} reports the temporal-split nRMSE on the
held-out M9 campaign across five real-data fractions. Synthetic
pre-training followed by fine-tuning outperforms training from
scratch at every fraction. As the real-data fraction decreases
from 5\% to 1\%, the nRMSE of the fine-tuned model remains close
to 2.3\%, whereas that of the from-scratch baseline increases
from 3.80\% to 6.42\%. The corresponding relative error reduction
increases from 38.6\% to 64.0\%.

\begin{table}[htbp]
\centering
\small
\begin{tabular}{lccc}
\toprule
\shortstack{Real-data\\fraction} & Real scratch & \shortstack{Pre-train +\\fine-tune} & Gain \\
\midrule
1\%   & $6.42 \pm 0.04$  & $\bm{2.31 \pm 0.17}$ & $\bm{+64.0\%}$ \\
5\%   & $3.80 \pm 0.12$  & $\bm{2.33 \pm 0.11}$ & $\bm{+38.6\%}$ \\
25\%  & $3.24 \pm 0.15$  & $\bm{2.25 \pm 0.10}$ & $\bm{+30.5\%}$ \\
50\%  & $2.79 \pm 0.17$  & $\bm{2.38 \pm 0.17}$ & $\bm{+14.9\%}$ \\
100\% & $2.49 \pm 0.17$  & $\bm{2.20 \pm 0.13}$ & $\bm{+11.4\%}$ \\
\bottomrule
\end{tabular}
\caption{Per-sample nRMSE (\%) on the temporal-split M9 test set (17{,}498 slices from 507 discharges) across five real-data fractions, comparing from-scratch training on real data (Real scratch) with synthetic pre-training plus real fine-tuning. Gain is the relative improvement over the from-scratch baseline, Gain $=(E_{\mathrm{scratch}} - E_{\mathrm{pretrained}})/E_{\mathrm{scratch}}\times100\%$. Results are mean $\pm$ std over three seeds. Bold marks the better (lower-error) value per row, and bold gains denote improvement over the baseline.}
 \label{tab:temporal_table}
\end{table}

\begin{figure}[t]
    \centering
    \includegraphics[width=0.5\columnwidth]{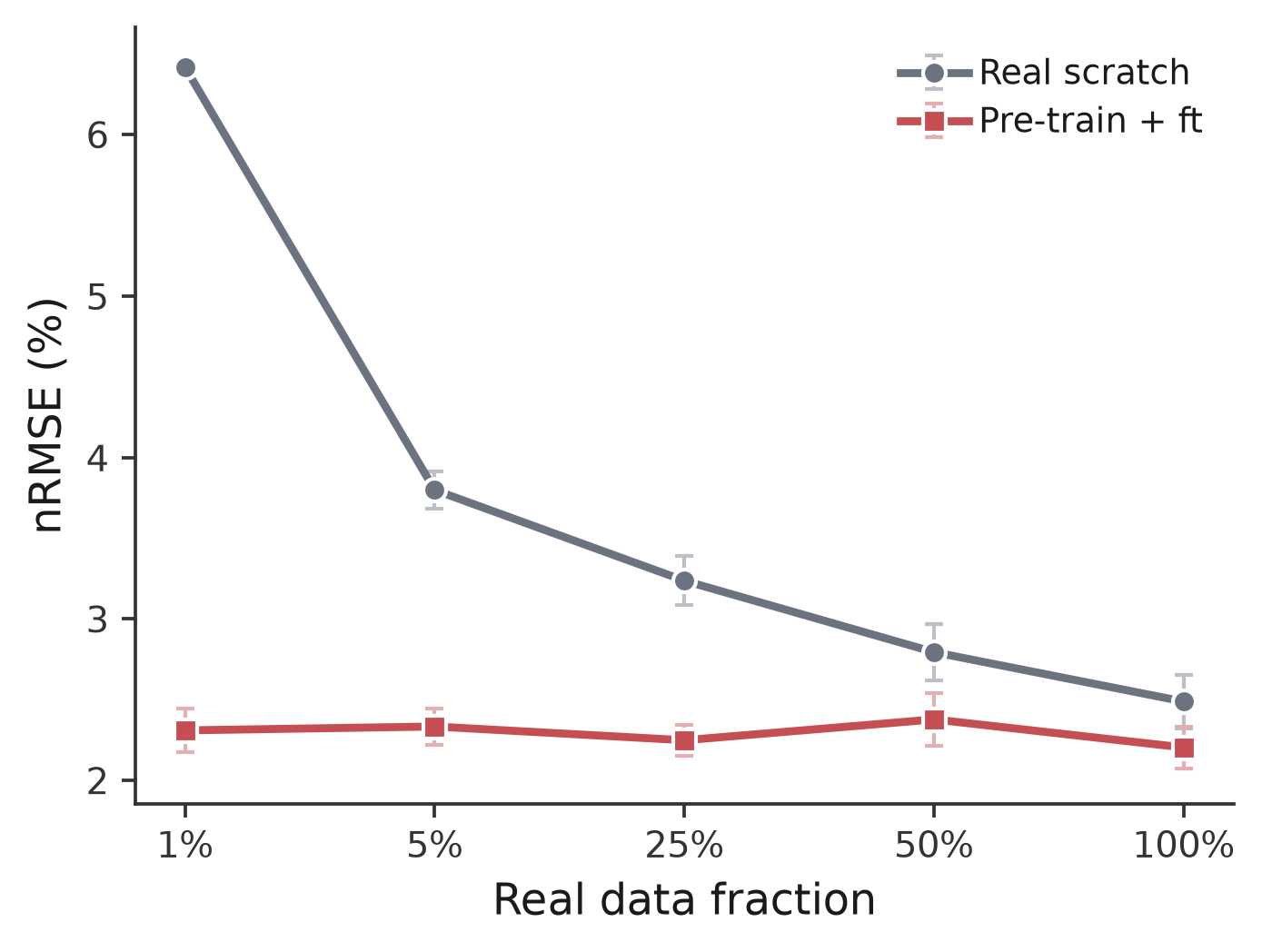}
    \caption{Per-sample nRMSE (\%) on the temporal-split M9 test set versus the real-data fraction, comparing from-scratch training on real data (Real scratch) with synthetic pre-training plus real fine-tuning. Pre-training improves accuracy at every fraction; the gap widens as real data become scarcer. Lower is better.}
    \label{fig:temporal_split_curves}
\end{figure}

\begin{figure}
    \centering
    \includegraphics[width=\linewidth]{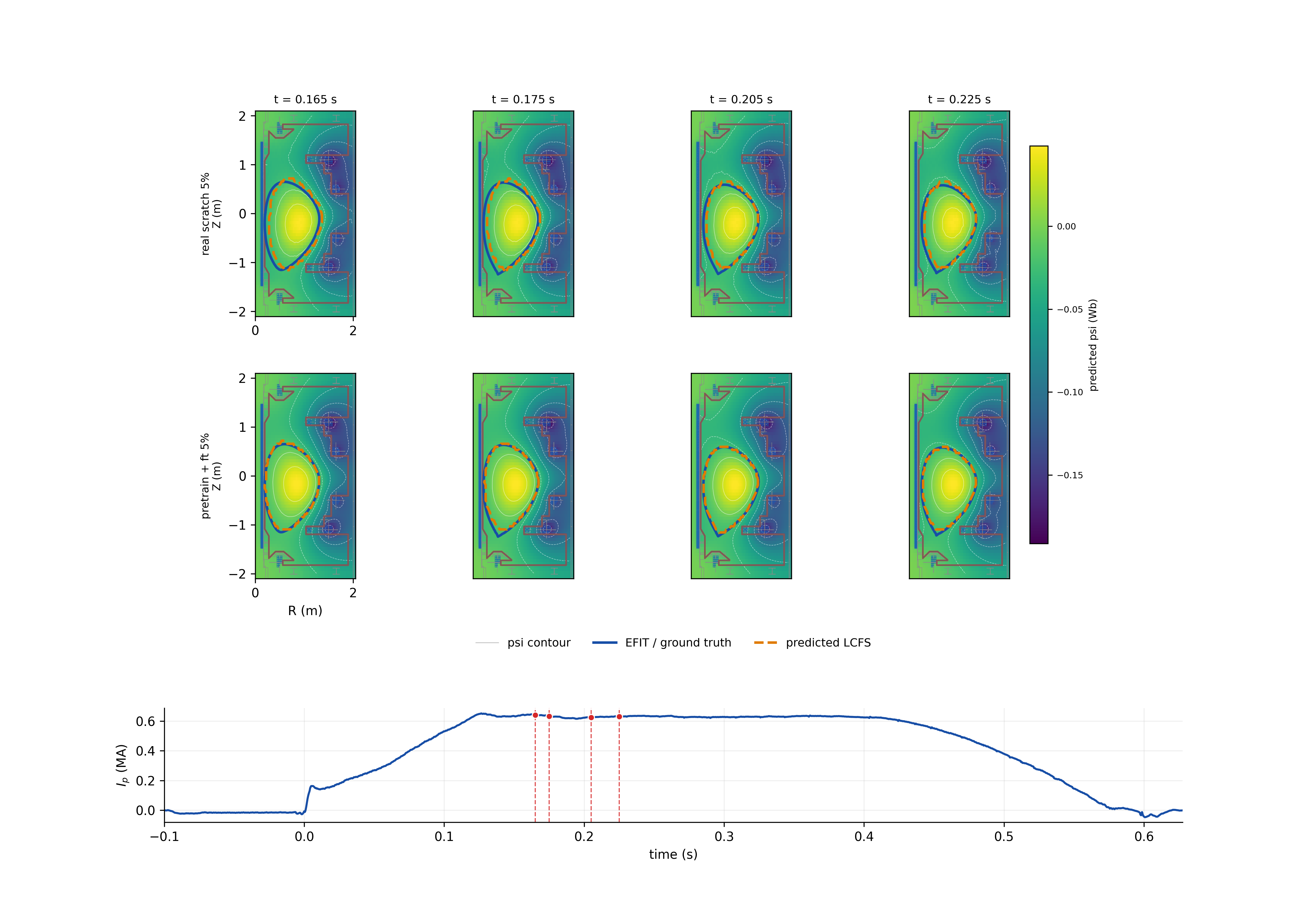}
    \caption{LCFS comparison of the from-scratch (top) and pre-trained (bottom) models on shot 29412 from the unseen M9 campaign at 5\% real data.}
    \label{fig:temporal_lcfs_5pct}
\end{figure}

\Cref{fig:temporal_lcfs_5pct} compares the LCFS predictions of the from-scratch and pre-trained models on shot 29412 from the unseen M9 campaign at 5\% real data: the pre-trained model closely reproduces the EFIT references, while the from-scratch model shows visible deviations.

\begin{figure}
    \centering
    \includegraphics[width=\linewidth]{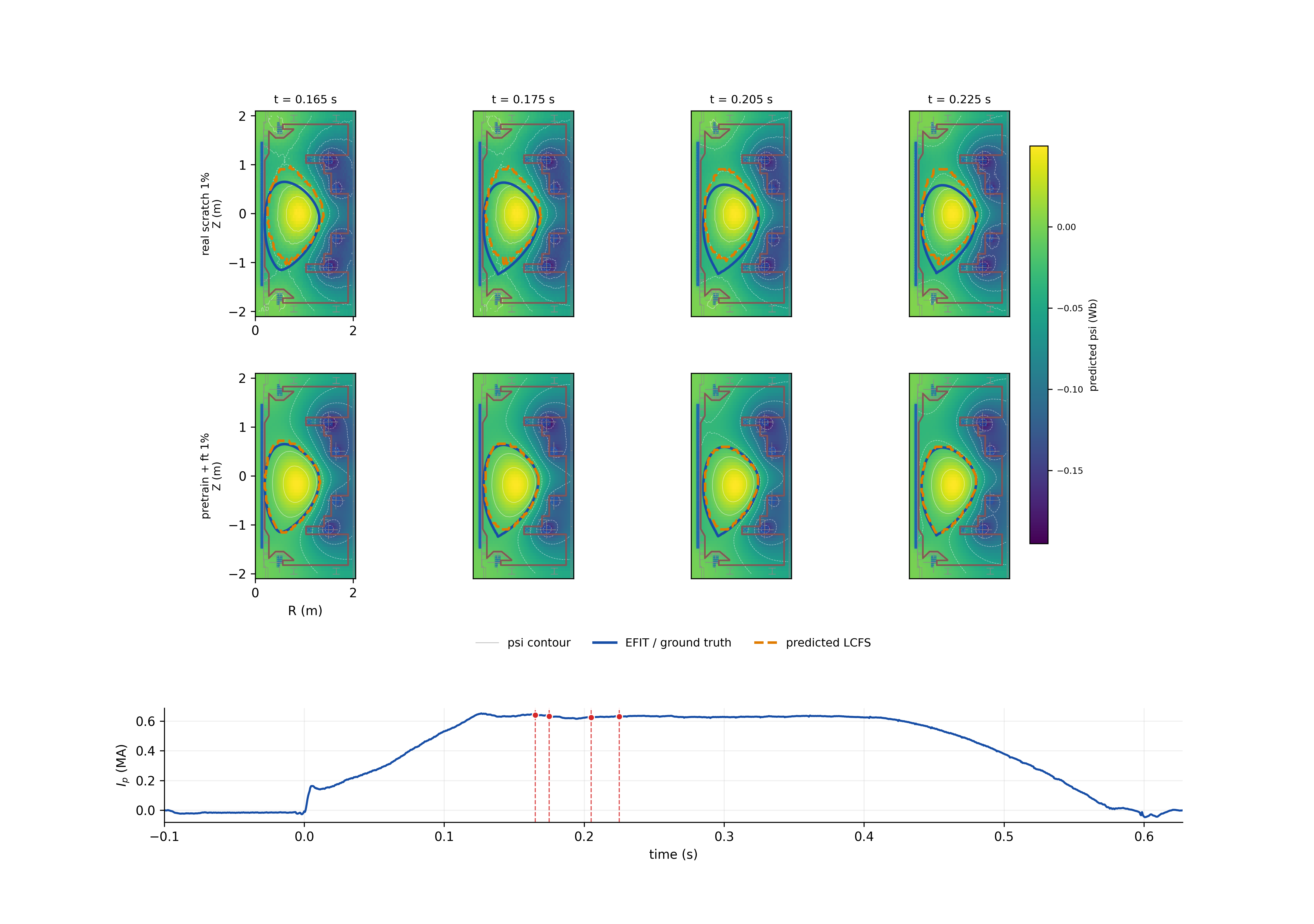}
    \caption{LCFS comparison of the from-scratch (top) and pre-trained (bottom) models on shot 29412 from the unseen M9 campaign at 1\% real data.}
    \label{fig:temporal_lcfs_1pct}
\end{figure}

At the more extreme 1\% real-data fraction of \Cref{fig:temporal_lcfs_1pct}, the from-scratch model suffers from severe boundary distortion, whereas the pre-trained model still maintains accurate LCFS predictions across all four time slices.

\subsection{Computational efficiency}
In terms of computational efficiency, our proposed model has an average processing time of approximately 0.036–0.039 ms per time slice on an NVIDIA GeForce RTX 5090 GPU, corresponding to a throughput of approximately 2.6–2.8 × 10\(^{4}\) slices s\(^{-1}\), meeting real-time control requirements. Compared with conventional iterative numerical equilibrium solvers (reported EFIT/EFIT++ runtimes are approximately \(0.1\)--\(2\,\mathrm{s}\) per time slice, depending on grid resolution and hardware~\citep{zwingmann2008integrated,appel2018equilibrium,havlicek2007magnetic}), the trained surrogate requires only a single forward pass during inference. Inference speed remains nearly unchanged across different real-data fractions and training strategies, demonstrating the low computational cost and potential for real-time equilibrium reconstruction.

\section{Conclusion}
\label{sec:conclusion}

We have demonstrated that physics-consistent synthetic equilibria, generated by FreeGSNKE forward modelling from real MAST discharges, provide an effective prior for equilibrium reconstruction. Because high-quality labels require expensive experimental shots and are scarce when a new campaign begins, we evaluated pre-training in settings that mirror real deployment, namely under in-distribution, out-of-distribution (shape), and cross-campaign (temporal) splits. The benefit is largest precisely where real data are scarce: when fine-tuned with only 1\% of the fully available real dataset, synthetic pre-training cuts reconstruction error by roughly 58\% under out-of-distribution shapes and 64\% under cross-campaign extrapolation relative to from-scratch training. The reconstructions are physically consistent: the predicted flux fields correlate with the EFIT references at $R\approx0.995$ and recover the total plasma current with $R\approx0.99$, with strong LCFS agreement. These results indicate that synthetic pre-training followed by real data fine-tuning can keep equilibrium reconstruction accurate when experimental labels are expensive or unavailable.

Several directions remain open. First, the synthetic pre-training framework could be validated across multiple tokamak devices to assess cross-machine transferability. Then, integrating uncertainty quantification into the reconstruction pipeline would enable confidence-aware equilibrium inference suitable for real-time control applications.

\ack{\textbf{AI use disclosure.} During the preparation of this work, the authors used DeepSeek-V4-Flash and OpenAI Codex (GPT-5.6 terrar) to assist with code debugging, grammatical correction of the manuscript, and the development of plotting and visualization code. The figures were plotted from the study data by running this code. The AI-assisted code was manually reviewed and verified. The authors take full responsibility for the accuracy, integrity, and originality of the manuscript and the reported results.}

\section{Data availability}
\label{sec:dataavail}
The code and results needed to reproduce this work are available at \url{https://github.com/Homingdung/mast-bridge}. The reconstruction backbone builds on the TokaMind transformer \cite{amorisco2026tokamind}. The raw data are from the open-access FAIR-MAST dataset \cite{jackson2024fairmast}.

\bibliography{ref}
\bibliographystyle{iopart-num}

\appendix
\section{Appendix}
\subsection{Statistics of FAIR-MAST dataset}
\Cref{tab:campaign_statistics} shows the statistics of the FAIR-MAST dataset by campaign.
\begin{table}[htbp]
\centering
\small
\begin{tabular}{lccc}
\toprule
Campaign & Shots & Slices & Fraction \\
\midrule
M5 & 992 & 23{,}433 & 10.95\% \\
M6 & 2{,}584 & 82{,}294 & 38.47\% \\
M7 & 2{,}470 & 81{,}565 & 38.13\% \\
M8 & 559 & 6{,}206 & 2.90\% \\
M9 & 507 & 17{,}498 & 8.18\% \\
Unknown & 174 & 2{,}928 & 1.37\% \\
\midrule
Total & 7{,}286 & 213{,}924 & 100\% \\
\bottomrule
\end{tabular}
\caption{Campaign distribution of the paired dataset: 213{,}924 slices from 7{,}286 discharges. M5+M6+M7 form the temporal-split training set (6{,}046 shots), M8 the validation set (559 shots), and M9 the test set (507 shots); 174 Unknown-campaign shots are excluded.}
\label{tab:campaign_statistics}
\end{table}

\subsection{Source terms and machine geometry}\label{app:source_terms}
\Cref{tab:gs_inputs} details the GS source terms and their data paths in the FAIR-MAST zarr structure. The perturbable inputs are $p',\,FF',\, I_{\mathrm{PF}}$; the machine geometry is kept fixed.

\begin{table}[htbp]
\centering
\small
\begin{tabular}{llll}
\toprule
Input & Data path (zarr) & Source & Role \\
\midrule
$p'$ & \code{equilibrium/dpressure\_dpsi} & EFIT reconstruction & source profile \\
$FF'$ & \code{equilibrium/f\_df\_dpsi} & EFIT reconstruction & source profile \\
$F_{\mathrm{vac}}$ & \code{bvac\_rmag} $\times$ \code{magnetic\_axis\_r} & EFIT reconstruction & $F_{\mathrm{vac}}=R_{\mathrm{mag}}B_{\mathrm{vac}}$ \\
$I_p$ & \code{magnetics/ip} & measured & current scale \\
$I_{\mathrm{PF}}$ & \code{pf\_active/coil\_current} & measured & vacuum field \\
$I_{\mathrm{passive}}$ & \code{pf\_passive/<family>\_current} & measured & vacuum field \\
\bottomrule
\end{tabular}
\caption{GS source terms, their zarr data paths, and their roles in the free-boundary solve.}
\label{tab:gs_inputs}
\end{table}

The static machine geometry is loaded from five \texttt{.pickle} files per shot and kept fixed in all solves, as detailed in \Cref{tab:machine_geometry}. Only \code{pf\_active}, \code{pf\_passive} and \code{limiter} affect the solve: the first two determine the vacuum field through the Green's functions, and the limiter bounds the current-carrying region and determines the plasma topology as either limited or diverted. \code{wall} and \code{magnetics\_probes} are used only for diagnostics and plotting.

\begin{table}[htbp]
\centering
\small
\begin{tabular}{ll}
\toprule
File & Content \\
\midrule
\code{MAST\_pf\_active.pickle} & PF coil geometry \\
\code{MAST\_pf\_passive.pickle} & passive structure geometry \\
\code{MAST\_limiter.pickle} & limiter geometry \\
\code{MAST\_wall.pickle} & wall geometry \\
\code{MAST\_magnetics\_probes.pickle} & magnetic probe geometry \\
\bottomrule
\end{tabular}
\caption{Machine geometry files and their influence on the free-boundary solve.}
\label{tab:machine_geometry}
\end{table}

\subsection{Data filtering}\label{app:data_filtering}
All slices are sampled every 5\,ms. The perturbation pool is the input to PCA perturbation and FreeGSNKE solving; accepted equilibria are those passing all seven quality criteria; the final dataset removes four slices from shot 13108 due to a permanently NaN magnetic probe channel.
\begin{table}[htbp]
\centering
\small
\begin{tabular}{lccl}
\toprule
Stage & Shots & Slices & Criterion \\
\midrule
Raw discharges & 11{,}573 & 1{,}343{,}995 & n/a \\
Clean pool & 11{,}148 & 818{,}139 & Diagnostics + EFIT + geometry \\
Flat-top window & 10{,}667 & 533{,}552 & $I_p \ge 0.85\,I_p^{\max}$ \\
Perturbation pool & 10{,}667 & 502{,}402 & Remove negative $I_p$, coil NaN, overlaps \\
FreeGSNKE solved & 10{,}037 & 476{,}954 & Solver convergence (tol $\le 10^{-8}$) \\
Accepted equilibria & 7{,}287 & 213{,}928 & Topology-anchored 7-criterion filter \\
Final paired dataset & 7{,}286 & 213{,}924 & Remove channel-missing samples \\
\bottomrule
\end{tabular}
\caption{Data preparation pipeline.}
\label{app:data_prep}
\end{table}

\subsection{Calibration of the diagnostic noise profile}
\label{app:noise_calibration}

The channel-specific noise scales were calibrated from residuals between real MAST measurements and diagnostic values forward-calculated from the corresponding EFIT equilibria. For calibration sample \(i\) and diagnostic channel \(c\), the residual is defined as
\begin{equation}
r_{i,c}
=
x^{\mathrm{obs}}_{i,c}
-
\mathcal{F}_c\left(\psi^{\mathrm{EFIT}}_i\right),
\label{eq:noise_residual}
\end{equation}
where \(x^{\mathrm{obs}}_{i,c}\) is the measured diagnostic value and \(\mathcal{F}_c(\psi^{\mathrm{EFIT}}_i)\) is its forward-model prediction. After removing non-finite residuals, the channel-specific noise standard deviation is estimated using the robust median absolute deviation:
\begin{equation}
m_c
=
\operatorname{median}_i(r_{i,c}),
\qquad
\sigma_c
=
1.4826\,
\operatorname{median}_i
\left|r_{i,c}-m_c\right|.
\label{eq:noise_calibration}
\end{equation}
The factor \(1.4826\) converts the median absolute deviation to the standard-deviation scale under a Gaussian distribution, while reducing sensitivity to outliers. The resulting \(\sigma_c\) values are fixed during synthetic-data generation and scaled by \(s\) according to \eqref{eq:noise}. The complete channel-wise profile used in all experiments is provided in the accompanying repository at \texttt{mast-bridge/configs/diagnostic\_features/noise\_profile.json}.

\subsection{Last-closed-flux-surface extraction}
\label{app:lcfs}
The network outputs the poloidal flux $\psi$ on a fixed $65 \times 65$ $(R,Z)$ grid, from which the plasma boundary is recovered without any explicit boundary label. The magnetic axis (O-point) is first located at the grid maximum of $\psi$. A level sweep is then performed: for a threshold
\begin{equation}
\ell = \psi_{\max} - f\,(\psi_{\max} - \psi_{\min}),
\label{eq:lcfs_level}
\end{equation}
with $f$ increasing from $0.05$ to $0.95$ in steps of $0.01$, the binary region $(\psi \ge \ell)$ is formed and its connected components are labeled. At each step we retain the component containing the O-point, rather than the largest component, since the predicted field can contain larger vacuum components that do not enclose the magnetic axis. The sweep terminates as soon as that component touches the computational-domain boundary (i.e., leaks into the vacuum), and the last non-leaking configuration is taken as the LCFS. The \texttt{skimage.measure.find\_contours} algorithm is then applied at level $0.5$, and the resulting contour is resampled uniformly to $170$ points by arc length to obtain the canonical representation. This adaptive threshold is necessary because the boundary flux value of the predicted field differs from that of EFIT; a fixed threshold would yield either a small contour near the axis or a region that has already leaked to the domain edge.

\subsection{Training details}
\label{app:training_details}
All experiments use the same Transformer architecture with $d_{\rm model}=64$, two Transformer layers, four attention heads, a feed-forward dimension of 128, and a dropout rate of 0.05.
\begin{table*}[htbp]
\centering
\small
\begingroup
\setlength{\tabcolsep}{4pt}
\renewcommand{\arraystretch}{1.20}

\begin{tabular*}{\textwidth}
{@{\extracolsep{\fill}}lcccccc@{}}
\toprule
& \multicolumn{3}{c}{\textbf{Random / Temporal}}
& \multicolumn{3}{c}{\textbf{Shape-OOD}} \\
\cmidrule(lr){2-4}
\cmidrule(lr){5-7}
\textbf{Parameter}
& \textbf{Pre-train}
& \textbf{Scratch}
& \textbf{Fine-tune}
& \textbf{Pre-train}
& \textbf{Scratch}
& \textbf{Fine-tune} \\
\midrule

Optimizer
& AdamW & AdamW & AdamW
& AdamW & AdamW & AdamW \\

Learning rate
& $10^{-4}$ & $10^{-4}$ & $10^{-4}$
& $2\times10^{-4}$ & $10^{-4}$ & $10^{-4}$ \\

Batch size
& 64 & 64 & 64
& 64 & 64 & 64 \\

Weight decay
& 0 & 0 & 0
& $10^{-4}$ & $10^{-4}$ & $10^{-4}$ \\

LR scheduler
& Cosine & Cosine & Cosine
& Cosine & Cosine & Cosine \\

Initialization
& Random & Random & Pre-trained
& Random & Random & Pre-trained \\

\bottomrule
\end{tabular*}
\endgroup

\caption{Training configurations for synthetic pre-training,
real-from-scratch training, and real-data fine-tuning under
the Random, Temporal, and Shape-OOD protocols.
Cosine denotes cosine learning-rate decay.}
\label{tab:training_configuration}
\end{table*}

For synthetic pre-training, the dataset is divided into training and validation subsets at the parent-discharge level, and the synthetic validation set is used for early stopping and checkpoint selection. No real MAST validation data are used during this stage. For real-data training and fine-tuning, the corresponding held-out real validation set defined in each evaluation protocol is used for model selection, while the test set is only used for final evaluation.

\end{document}